\documentclass[aip,pof,natbib]{revtex4-1}

\usepackage{graphicx}% Include figure files
\usepackage{dcolumn}% Align table columns on decimal point
\usepackage{bm}% bold math
\usepackage[utf8]{inputenc}
\usepackage[T1]{fontenc}
\usepackage{mathptmx}
\usepackage{etoolbox}
\usepackage{xcolor}
\usepackage{url}
\usepackage[normalem]{ulem}
\usepackage{subfigure}
\usepackage{amsmath}
\usepackage{epstopdf, epsfig}

\makeatletter
\def\@email#1#2{%
 \endgroup
 \patchcmd{\titleblock@produce}
  {\frontmatter@RRAPformat}
  {\frontmatter@RRAPformat{\produce@RRAP{*#1\href{mailto:#2}{#2}}}\frontmatter@RRAPformat}
  {}{}
}%
\makeatother
\begin{document}

%\preprint{AIP/123-QED}

\title[Role of varying Reynolds number for flow past a rotating cylinder at high rotation rate]{Role of varying Reynolds number for flow past a rotating cylinder at high rotation rate\\}

% Force line breaks with \\
\author{Aditi Sengupta \footnote{corresponding author}}
\email{aditi@iitism.ac.in}

\author{Santosh Kumar}%

\author{Sanjeev Kumar}

\affiliation{Department of Mechanical Engineering, Indian Institute of Technology Dhanbad, Jharkhand-826 004, India.
%\\This line break forced with \textbackslash\textbackslash
}%

%\affiliation{Department of Mechanical Engineering, Indian Institute of Technology Dhanbad, Jharkhand-826 004, India.}

\date{\today}% It is always \today, today,

\begin{abstract}
The present study reports comprehensive bifurcation analysis of flow past a rotating cylinder at a fixed rotation rate by varying free-stream Reynolds number ($Re_{\infty}$) from 1000-6000 in intervals of 50. Two-dimensional compressible Navier-Stokes equations are solved using dispersion relation preserving numerical methods over 101 test cases, amounting to $10^6$ core hours of computing.  The dataset produced from high-fidelity simulations serve as useful benchmarking tools for testing compressible flow solvers, estimating unsteady force distribution and vorticity dynamics. For moderate $Re_{\infty}$, rotation induces circulation that reduces pressure drag with increasing $Re_{\infty}$. For higher $Re_{\infty}$, boundary layer becomes thinner with suppressed flow separation, but effect of rotation saturates. Thus, benefits of increasing $Re_{\infty}$ taper off and pressure recovery stalls. The bifurcation analysis reveals a critical $Re_{\infty}$ of 5650 beyond which global behavior of Magnus-Robins effect changes significantly. Supercritical flow is receptive to time-dependent instabilities and structures in wake of the cylinder become dynamically unstable. Even small changes in $Re_{\infty}$ leads to different instantaneous force distributions and sharp fluctuations in lift and drag calculations. Stronger, coherent vortices in the wake generate consistent, high-energy periodic signals, contributing to strong Fourier amplitudes in spectra. An artificial neural network (ANN) is trained using simulation datasets to serve as fast, inexpensive alternatives for calculating lift, drag, and onset time of instability. The ANN reduces time required for simulation by 99.9\%, enabling dense parametric sweeps. Maximum accuracy achieved for the ANN is between 90-99\% for the parameters examined. 
\end{abstract}

\maketitle

\section{\label{sec:level1}Introduction}

The flow past a rotating circular cylinder has long served as a canonical problem in fluid mechanics, revealing rich and complex dynamics that depend on Reynolds number, rotation rate, and, in compressible regimes, the Mach number \cite{white1994fluid}. While the incompressible case has been extensively studied due to its applications in aerodynamics, rotating machinery, and vortex-induced vibrations \cite{mittal2003flow, sengupta2003temporal, stojkovic2002effect, glauert1957flow}, compressibility introduces additional layers of complexity, including shocklets, pressure waves, and modified instability characteristics \cite{teymourtash2017compressibility, sundaram2021multiscale, suman2022novel}. This problem has wide ranging applications in aerodynamics, in the design of centrifuges, flow control, sports, ballistics, and meteorology. The flow past a rotating cylinder is often used by researchers to explain the mechanism for lift generation, popularly known as the Magnus–Robins effect \citep{prandtl1926application}. Three parameters influence the dynamics of the problem depicted in Fig. \ref{fig1}: the speed of the oncoming flow $U_{\infty}$, and thereby the reference Reynolds number, $Re_{\infty}$, and the nondimensional surface speed of rotation $U_s/U_{\infty}$. Here, the subscript $\infty$ refers to the free stream conditions. A rotating cylinder experiences pressure differences on its surface due to the velocity induced by the rotation, producing lift. From a theoretical perspective, this problem has produced several investigations \cite{tokumaru1993lift} which showed that the Prandtl’s hypothesis about the maximum circulation limit is invalid. For incompressible flows, the instability mechanisms for shedding \cite{pralits2010instability} and temporal instabilities at high rotation rates have been explored \cite{sengupta2003temporal}. Pralits {\it et al.} \cite{pralits2010instability} have studied the linear instability of two-dimensional flow past a rotating cylinder to analyze the distinct instability mechanisms for the appearance of von Karman and single-sided vortex shedding modes. The authors report that the von Karman mode disappears when $U_s/U_{\infty}\approx 2$ and the single-sided shedding mode appears when $4.85 \le U_s/U_{\infty} \le 5.17$.

Previous studies have considered the instability during the Magnus–Robins effect in an incompressible flow. However, compressibility effects play a major role when the speed of oncoming flow is high and/or at high rotation rates \cite{salimipour2019surface, suman2022novel}. These studies
have shown that significant compressibility effects including unsteady
shock wave–boundary layer interactions occur for flows with high
rotation rates even when the free-stream Mach number, $M_{\infty}$ is low. The maximum surface speed of rotation considered by Suman {\it et al.} \cite{suman2022novel} is $U_s = 14U_{\infty}$, where supersonic pockets and weak shocks were present in the flow. In addition to the observed compressibility effects, significant heating of the fluid was reported, with maximum temperatures reaching values which are 3.8 times the free stream temperature. A marginal moderation/limiting of the lift force is reported due to compressibility \cite{teymourtash2017compressibility}. Nagata {\it et al.} \cite{nagata2018direct} observed a similar trend in lift distribution for the compressible flow past a rotating sphere. A rotating cylinder in compressible flow generates a lift force through circulation, an effect first explained by the Kutta–Joukowski theorem in the inviscid limit \cite{kundu2012fluid}. At low-to-moderate Mach numbers, the interaction between rotation-induced vorticity and compressibility can either stabilize or destabilize the flow depending on the rotation rate and boundary-layer dynamics \cite{nishioka1978mechanism}. It has been identified that increasing the rotation rate inhibits vortex shedding which is completely suppressed for a critical rotation rate \cite{kang1999laminar}. This critical rotation rate is found to have a logarithmic dependence on the $Re_{\infty}$. The mean lift increased linearly and mean drag decreased with an increase in rotation rate, in contrast to potential flow theory. For sufficiently high rotation rates, the flow can transition from a periodic vortex street to steady, asymmetric patterns or even re-laminarized wakes \cite{tokumaru1991rotary, mittal2003flow}. Experiments on flow past two uniformly rotating cylinders in a side-by-side configuration \cite{kumar2011flow}, explored two scenarios: cylinders moving upstream in one case (inward rotation) and downstream in the other (outward rotation). Vortex shedding is suppressed for inward rotation while for outward rotation, vortex shedding suppression depended on $Re_{\infty}$ and the gap between the cylinders. In later computations of flow around two co-rotating cylinders \cite{darvishyadegari2019heat}, co-rotating the cylinders showed additional events such as azimuthal displacement of front stagnation points and development of negative lift coefficient for both cylinders.

Compressibility and rotation rates have an important role in the compressible flow past a rotating cylinder. As $M_{\infty}$ increases from nearly-incompressible to subsonic values, a study \cite{van2007compressible} identified $M_{\infty} \approx 0.3$ as a threshold where compressibility visibly alters boundary-layer/wake behavior. Above this, density variations and acoustic coupling modify instability growth rates and vortex roll-up. Several recent numerical studies \cite{liu2023numerical, xue2024compressibility} reported that moderate $M_{\infty}$ tend to weaken coherent vortex roll-up and reduce shedding amplitude, i.e. they act in a stabilizing manner on the classic 2D shedding mode for certain Re ranges. This effectively shifts the classical critical $Re_{\infty}$ for coherent shedding. The authors \cite{liu2023numerical, xue2024compressibility, salimipour2019surface} reported compressibility-induced changes in shedding strength, Strouhal number and wake stability. However, compressibility does not always stabilize: at some parameter combinations compressibility can introduce or amplify other (acoustic or shear-layer) instabilities, especially as $M_{\infty}$ approaches transonic values or when shocklets/strong pressure gradients appear. Biringen {\it et al.} \cite{biringen1993numerical} noted that compressibility can stabilize or destabilize depending on rotational speeds, gap widths, and three-dimensional effects, for counter-rotating cylinders. Rotation, on the other hand, changes the surface velocity and the mean vorticity distribution. The co-rotating side receives added momentum which delays separation while the counter-rotating side loses momentum and separates earlier. As a result the global eigenmode responsible for the von Kármán street can be stabilized beyond a rotation threshold. Mittal and Kumar \cite{mittal2003flow} reported a critical rotation for disappearance of shedding near $U_s = 1.9 U_{\infty}$ for $Re_{\infty} = 200$. The critical rotation rate depends on $Re_{\infty}$, with some some authors observing re-emergence of secondary periodicity \cite{sundaram2021multiscale, sengupta2003temporal, kumar2011flow} or new modes for certain $Re_{\infty}$ ranges, so the suppression is not strictly monotonic across all $Re_{\infty}$ and very high rotation can introduce new instabilities \cite{suman2022novel}.

Bifurcation theory provides a systematic framework to classify the transitions and identify the critical parameters for which the system's qualitative behavior changes. In incompressible flow, Hopf and pitchfork bifurcations have been reported near critical Re and rotation rates \citep{blackburn1999study}. Tokumaru and Dimotakis \cite{tokumaru1991rotary} reported experimental investigation of incompressible flow past a rotating cylinder for $Re_{\infty}$ in the range of 3800-6800 and for surface speed, $U_s/U_{\infty}$ in range of  0.5 to 10. The mean lift coefficient was investigated in detail for steady and oscillatory rotations, showing a marked departure from Prandtl's limit. Kang {\it et al.} \cite{kang1999laminar} performed two-dimensional incompressible flow simulations for $Re_{\infty} = 60, 100$, and 160 and for surface speed, $U_s/U_{\infty}$ in range of  0 to 2.5. The purpose of the work was to facilitate control of vortex shedding and understanding the underlying flow mechanism. A critical rotational speed was established beyond which shedding disappears. Stojkovic {\it et al.} \cite{stojkovic2002effect} also conducted incompressible flow computations for $Re_{\infty} = 60 - 200$ and for surface speed, $U_s/U_{\infty}$ in range of  0 to 6. The behavior of a new vortex shedding mode was investigated, showing the existence of the second shedding mode for the entire $Re_{\infty}$ range. A complete bifurcation diagram was compiled for the shedding modes. The computations of Pralits {\it et al.} \cite{pralits2010instability} for incompressible flow with $Re_{\infty} = 100$ and for surface speed, $U_s/U_{\infty}$ in range of  0 to 7, reported a global linear instability analysis. Instability mechanisms for the first and second shedding modes were analyzed using structural sensitivity and perturbation kinetic energy budget. Kumar {\it et al.} \cite{kumar2011flow}'s experiments for incompressible flow with $Re_{\infty} = 200, 300$, and 400, and for surface speed, $U_s/U_{\infty}$ in range of  0 to 5, provided a global view of the wake structure. Single-sided vortex shedding was presented for the first time in experiments. Pralits {\it et al.} \cite{pralits2013three} performed a linear stability analysis in an incompressible formulation to investigate two-dimensional and three-dimensional onset of bifurcation. Discrepancies between numerical and experimental findings at large rotation rates were explained. The incompressible simulations of Akoury {\it et al.} \cite{el2008three} with $Re_{\infty} \le 500$, and for surface speed, $U_s/U_{\infty} < 5.5$, explored the transition at low $Re_{\infty}$. Rotation attenuated the secondary instability and increased the critical Reynolds number. In the review of Rao {\it et al.} \cite{rao2015review} incompressible flow with $Re_{\infty} \le 400$, and for surface speed, $U_s/U_{\infty} \le 7$, two- and three-dimensional transitions occurring with increasing rotation rates are detailed. Floquet analysis showed the presence of five three-dimensional modes.

However, the bifurcation structure of compressible flows, particularly at high rotation rates, remains far less understood. Teymourtash and Salimipour \cite{teymourtash2017compressibility} explored in their compressible computations with $M_{\infty}$ in the range 0.05 to 0.4, $Re_{\infty}= 40, 60, 120, 200$, and for surface speed, $U_s/U_{\infty} \le 14$, the compressibility effects on shedding, lift and drag coefficients. At high $M_{\infty}$, maximum mean lift coefficient, mean drag coefficient, and Strouhal number became independent of $Re_{\infty}$. Salimipour and Anbarsooz \cite{salimipour2019surface} investigated the effects of surface temperature on the compressible flow past a rotating cylinder with $M_{\infty}$ in the range 0.1 to 0.4, $Re_{\infty}= 200$, and for surface speed, $U_s/U_{\infty} \le 7$. Increasing the surface temperature reduced the lift coefficient considerably and a new mode of vortex shedding was reported. Sundaram {\it et al.} \cite{sundaram2021multiscale} performed highly accurate compressible simulations with $M_{\infty} = 0.05, 0.1, 0.14$, $Re_{\infty}= 3800$, and for surface speed, $2 \le U_s/U_{\infty} \le 6$. The effect of compressibility was studied and a detailed force distribution and vortex dynamics of the multi-scale temporal instabilities were reported for the first time. For a fixed $M_{\infty} = 0.1$, and $Re_{\infty} = 3800$, Suman {\it et al.} \cite{suman2022novel} reported compressible simulations with high dimensionless rotation rates of $U_s/U_{\infty} = 12$ and 14. The temporal instability was examined using a compressible enstrophy transport equation and mechanisms contributing to it were identified. Compressibility not only alters the instability thresholds but also couples density and pressure fluctuations with vorticity dynamics. This potentially gives rise to novel bifurcation pathways \citep{theofilis2003advances}, which has not received a systematic parametric investigation so far. 

The present study performs a detailed bifurcation analysis of two-dimensional compressible flow past a rotating cylinder at a fixed dimensionless rotation rate of $U_s/U_{\infty} = 10$, for the first time. As the cylinder is rotating at a high speed, the surrounding fluid is entrained to the surface with significantly high speed, which may not be adequately captured by an incompressible formulation. Thus, the present research, conducted for a fixed free-stream $M_{\infty}$ of 0.1, requires an accurate compressible formulation to understand the role of compressibility in evoking the temporally unstable Magnus-Robins effect. Emphasis is placed on identifying the role of free-stream Reynolds number in wake transition, and the parameter regions associated with symmetry breaking, limit cycles, and quasi-periodicity. The findings aim to extend the current understanding of vortex dynamics in high-speed rotational flows and provide a foundation for flow control strategies in aero-propulsive systems. Additionally, the time-resolved database involving 101 two-dimensional compressible flow simulations over $10^6$ core hours serves as a benchmark for testing compressible flow solvers and for the development and training of supervised machine learning techniques.

\section{Problem formulation of the rotating cylinder}

\begin{figure*}%% placement specifier
\centering
\includegraphics[width=.9\textwidth]{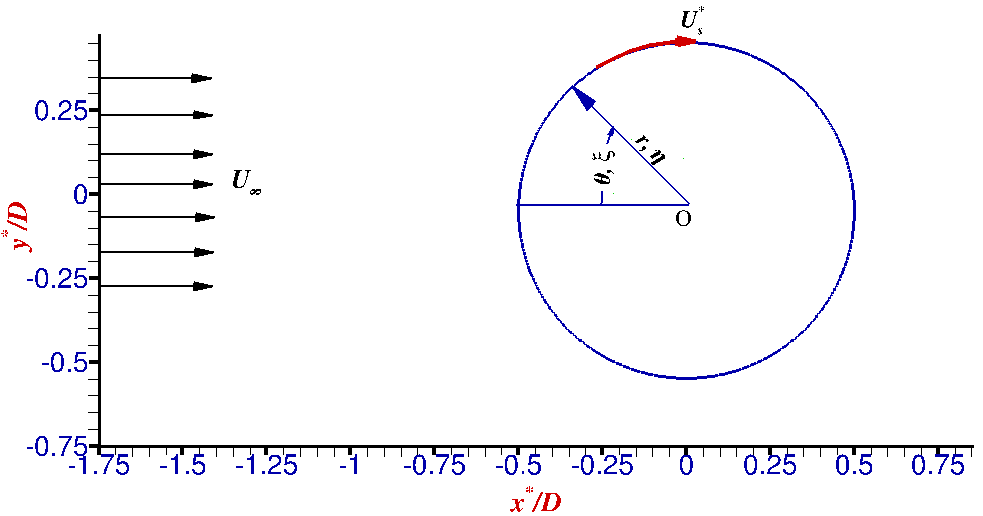}
%% Use \caption command for figure caption and label.
\caption{Schematic of uniform flow past a rotating cylinder with surface speed, $U^*_s$.}
\label{fig1}
\end{figure*}

The schematic for two-dimensional flow past a rotating cylinder in free stream is given in Fig. \ref{fig1}. The flow field is generated by an infinitely long cylinder of diameter, $D$ which is rotating about its axis with angular velocity, $\Omega^*$.  The uniform flow having velocity, $U_{\infty}$ is translating from left to right and impinging the rotating cylinder at right angles to its axis. The azimuthal angle ($\theta$) is marked with respect to the most upstream point of the cylinder on the windward side. Here, the rotation is indicated by the tangential surface speed, $U^*_s$, evaluated as $U^*_s = \Omega^* D/2$. On the surface of the cylinder we prescribe no-slip and adiabatic boundary conditions. Thermal boundary conditions are implemented in a direction normal to the surface of the cylinder. The far-field is taken at 30$D$ where characteristic based \citep{pulliam1986implicit} boundary condition is used.

We solve the two-dimensional compressible Navier-Stokes equations, formulated by established notations \citep{hoffmann2000computational}, as follows,

\begin{eqnarray}
\frac{\partial {\hat{Q}}}{\partial t} + \frac{\partial {\hat{E}}}{\partial x} + \frac{\partial {\hat{F}}}{\partial y} = \frac{\partial {\hat{E_{v}}}}{\partial x} + \frac{\partial {\hat{F_{v}}}}{\partial y}
\label{eq1} 
\end{eqnarray}

\noindent where the conservative variables are given as, $\hat{Q} = [ \rho \;\; \rho u \;\; \rho v \;\; e_t]^T$. The convective flux vectors are similarly given as,

\begin{equation}
\hat{E} = [ \rho u \;\; \rho u^2 +p \;\; \rho uv \;\; (\rho e_t + p) u]^T 
\label{eq2} 
\end{equation}
\begin{equation}
\hat{F} = [ \rho v \;\; \rho uv \;\; \rho v^2 +p \;\; (\rho e_t + p) v]^T
\label{eq3} 
\end{equation}

\noindent and the viscous flux vectors are given as,

\begin{equation}
\hat{E_{v}}= [ 0\;\; \tau _{xx} \;\; \tau _{xy} \;\; (u \tau _{xx} +v \tau _{xy} -q _{x})]^T
\label{eq4} 
\end{equation}
\begin{equation}
\hat{F_{v}}= [0 \;\; \tau _{yx} \;\; \tau _{yy} \;\; (u \tau _{yx} +v \tau _{yy} -q _{y})]^T
\label{eq5} 
\end{equation}

In Eqs. \eqref{eq1} to \eqref{eq5}, $\rho$, $u$, $v$, $e_{t}$, and $p$ denote dimensionless values of density, velocity components, total specific energy, and pressure, respectively. These physical variables are normalized with respect to the free-stream density ($\rho_{\infty}$), free-stream velocity ($U_{\infty}$), free-stream temperature ($T_{\infty}$), free-stream dynamic viscosity ($\mu_{\infty}$), length scale ($D$), and the time scale, ($D/U_{\infty}$). The dimensionless parameters, namely the Prandtl number ($Pr$), free-stream Reynolds number ($Re_{\infty}$), and free-stream Mach number, ($M_{\infty}$), are defined as follows:

\begin{equation*}
Pr = \frac{\mu_{\infty} C_{p}}{\kappa}	;\ Re_{\infty} = \frac{\rho_{\infty} U_{\infty} D}{\mu_{\infty}}	;\ M_{\infty} = \frac{U_{\infty}}{\sqrt{\gamma R^* T_{\infty}}}
\end{equation*}    

\noindent where $\gamma=1.4$ represents the ratio of specific heat capacity at constant pressure ($C_p$) to constant volume ($C_v$). The pressure scale ($p_{\infty}$) is evaluated as $p_{\infty} = \rho_{\infty} R^* T_{\infty}$, where $R^*$ is the dimensional gas constant, non-dimensionalized as $R = R^*T_{\infty}/U^2_{\infty}$. The equation of state for an ideal gas, $p = \rho R T$ is used to relate the state variables and $e_t$ is defined as $e_t = C_v T + \frac{1}{2} (u^2 + v^2)$. Heat conduction terms are given by

\begin{equation*}
q_{x}= - \frac{\mu}{Pr Re_{\infty} (\gamma -1) M_{\infty}^2} \frac{\partial {T}}{\partial x}; \; \;\\
q_{y}= - \frac{\mu}{Pr Re_{\infty} (\gamma -1) M_{\infty}^2} \frac{\partial {T}}{\partial y}
\end{equation*}

\noindent The components of the Newtonian viscous stress tensors, $\tau_{xx},\tau_{xy},\tau_{yx},\tau_{yy}$, are defined as 

\begin{equation*}
\tau_{xx} = \frac{1}{Re_{\infty}}\Big[ 2\mu \frac{\partial u}{\partial x} + \lambda \nabla \cdot \vec{V} \Big]; \; \\
\tau_{yy} = \frac{1}{Re_{\infty}}\Big[2\mu\frac{\partial v}{\partial y}+ \lambda \nabla \cdot \vec{V} \Big]; \; \\
\tau_{xy} = \tau_{yx} = \frac{\mu}{Re_{\infty}}\Big[\frac{\partial u}{\partial y} + \frac{\partial v}{\partial x} \Big]
\end{equation*}

\noindent Here, Stokes' hypothesis ($\lambda = -\frac{2}{3} \mu $) is used in calculating stress tensor. To calculate viscosity as a function of temperature, Sutherland's law is applied. The equations from the Cartesian space ($x$, $y$) are transformed to body-fitted computational grid ($\xi, \eta$) using the following relations: $\xi = \xi (x, y)$ and $\eta = \eta (x, y)$. The transformed plane equations in strong conservation form are given as, 

\begin{eqnarray}
\frac{\partial {{\bf Q}}}{\partial t} + \frac{\partial {\bf E}}{\partial \xi} + \frac{\partial {\bf F}}{\partial \eta} = \frac{\partial {\bf E_{v}}}{\partial \xi} + \frac{\partial {\bf F_{v}}}{\partial \eta}
\label{eq6} 
\end{eqnarray}   

\noindent with the state variables and flux vectors, given as 
\begin{equation*}
{\bf Q} = \hat{Q}/J; \; \; \; \\
{\bf E} = (\xi _{x} \hat{E} +\xi _{y} \hat{F})/J; \; \; \; \\
{\bf F} = (\eta _{x} \hat{E} +\eta _{y} \hat{F})/J; \; \; \;\\
{\bf E_{v}} = (\xi _{x} \hat{E_{v}} +\xi _{y} \hat{F_{v}})/J; \; \; \;\\
{\bf F_{v}} = (\eta _{x} \hat{E_{v}} +\eta _{y} \hat{F_{v}})/J
\end{equation*}

\noindent and $J$ is the Jacobian of the grid transformation given by $
J=\frac{1}{x_{\xi} y_{\eta} -x_{\eta} y_{\xi}}$. The grid metrics, $\xi_{x}$, $\xi_{y}$, $\eta_{x}$, and $\eta_{y}$, are computed during the creation of an O-type grid using a hyperbolic grid generation technique in Pointwise. These are expressed as follows: $\xi_{x} = J y_{\eta}$; $\xi_{y} = -J x_{\eta}$; $\eta_{x} = -J y_{\xi}$; $\eta_{y} = J x_{\xi}$ and are used in calculating partial derivatives of the Cartesian grid in the transformed plane. For the computational domain shown in Fig. \ref{fig1}, an $O$-type grid consisting of 401 equidistant points in $\xi$-direction and 450 points in $\eta$-direction is adopted. In the $\eta$-direction, grid points are clustered near the wall using the following tangent hyperbolic stretching function: $$r(\eta) = 0.5 + r_{max} \biggl[ 1-\frac{tanh\beta \biggl (\frac{N_{\eta}-\eta}{N_{\eta}-1}\biggr)}{tanh \beta}\biggr]$$ where $N_{\eta}$ is number of points in $\eta$-direction, $r_{max}$ is determined from far-field boundary (set to 30$D$ in present simulations), and $\beta$ is stretching parameter which controls wall spacing, $\Delta r_{wall}$. Previous work reported on rotating cylinder \citep{sundaram2021multiscale} performed a mesh independence study using three wall resolutions, $\Delta r_{wall}$ = 0.0005$D$, 0.001$D$ and 0.002$D$. For the first two wall spacings, the coefficient of lift ($C_l$) was found to be identical. Hence, for the test cases reported here, we use the finest resolution, i.e. $\Delta r_{wall} = 0.0005D$ for which $\beta = 3$.

\subsection{Numerical methodology}
To solve the governing equation in Eq. \eqref{eq6}, one requires spatial discretization of the convective flux terms. This is done using an optimized upwind compact scheme, OUCS3, which reports near-spectral accuracy for a wider range of wavenumbers than other numerical methods \citep{sagaut2023global}. The viscous flux derivatives are evaluated using a non-uniform explicit central difference method. Fourth-order diffusion, designed by Jameson \cite{jameson2017origins}, is added to eliminate high wavenumber numerical noise with a coefficient of 0.015. The time advancement is performed with an optimized three-stage Runge-Kutta scheme (OCRK3) \citep{sagaut2023global} with a non-dimensional time-step, $\Delta t = 1.25 \times 10^{-5}$. The combined time-space discretization has been analysed using global spectral analysis, reporting preservation of the dispersion relation between spatial and temporal scales in the flow. The in-house, finite-difference based Navier-Stokes' solver has been parallelized using message passage interface. The non-overlapping high accuracy parallel (NOHAP) algorithm \citep{sundaram2023non} has been adopted to completely eliminate errors at sub-domain boundaries between successive processors without the need for large number of overlapping points. This parallel algorithm has been benchmarked for a variety of incompressible and compressible problems, showing identical results as sequential computing. 

\begin{figure*}%% placement specifier
\centering
\includegraphics[width=.7\textwidth ]{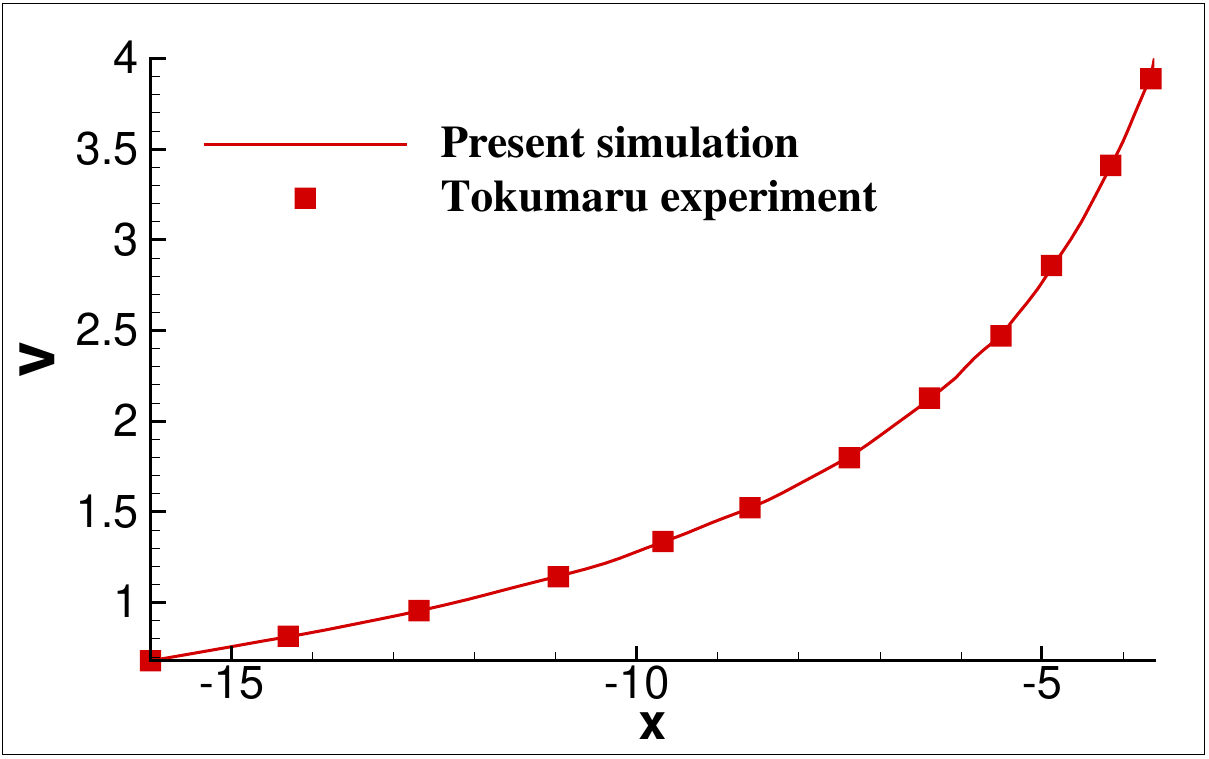}
%% Use \caption command for figure caption and label.
\caption{Comparison of computed and experimental \cite{tokumaru1991rotary} normalized transverse velocity upstream of cylinder with steady rotation at $Re_{\infty}$ = 3800 and $U^*_s = 10 U_{\infty}$. The free-stream Mach number, $M_{\infty}$ of the simulation is fixed at 0.1 to mimic the experimental setup.}
\label{fig2}
\end{figure*}

The present work aims to perform a bifurcation analysis for flow past a rotating cylinder at a high dimensionless rotation rate for which the surface speed is taken as $U^*_s = 10 U_{\infty}$. The free-stream Reynolds number, $Re_{\infty}$ for the incoming flow is varied from 1000 to 6000 in intervals of 50, resulting in 101 two-dimensional compressible flow simulations. The free-stream Mach number, $M_{\infty}$ is fixed at 0.1, so that one is operating within the incompressibility limit. The $Pr$ is fixed at 0.71 for air and the temperature scale is taken as $T_{\infty} = 288.15$K for which the density scale is $\rho_{\infty} = 1.2256kg/m^3$. The present numerical setup has been compared with experiments of Tokumaru and Dimotakis \cite{tokumaru1993lift} in a previous work \citep{sundaram2021multiscale}. In the experiment \cite{tokumaru1991rotary}, a cylinder exhibiting steady rotation and rotary oscillations was considered for two $Re_{\infty}$ = 3800 and 6800, for dimensionless rotation rates ranging from 0.5 to 10. In Fig. \ref{fig2}, we compare our compressible solver computations of normalized transverse velocity ($v$) with that of the experiment for steady rotation at $Re_{\infty}$ = 3800 and $U^*_s = 10 U_{\infty}$. The normalized transverse velocity is measured at probe locations upstream of the cylinder. The experimental measurements are shown to have a good agreement with the computed transverse velocity, which demonstrates that the numerical framework adopted here is sufficient to capture the essential flow physics.

Each test case is computed using 40 cores over 240 compute hours till limit cycle oscillation stage is reached in the amplitude of the lift and drag coefficients. For evaluating 101 test cases, the total computational time required is approximately 1 million core hours. This has been done primarily to produce a benchmark database for high-fidelity simulation of flow past a rotating cylinder at a high dimensionless rotation rate, subject to varying $Re_{\infty}$. Furthermore, an artificial neural network (ANN) is designed which can model three critical parameters related to the Magnus-Robins effect in flow past a rotating cylinder. This serves as a substitute to expensive computations.

\section{Results and discussion}
In this section, we will explore the vorticity dynamics of the flow past a rotating cylinder subject to different free-stream $Re_{\infty}$. The dominant temporal scales for different $Re_{\infty}$ are captured by performing a Fast Fourier transform (FFT) of the vorticity time-series. The unsteady force distribution over the rotating cylinder will be compared for varying $Re_{\infty}$ by tracking the coefficients of drag and lift. The onset of the temporal instability as a function of $Re_{\infty}$ will be explored and the system's bifurcation will be explained in detail. Finally, a data-driven model of the critical parameters will be provided by application of an ANN. This captures the behavior of the system for benchmarking and further analysis without the need for expensive, high fidelity simulations.

\subsection{Vorticity dynamics at high dimensionless rotation rate for different Reynolds number}

\begin{figure*}%% placement specifier
\centering
\includegraphics[width=\textwidth ]{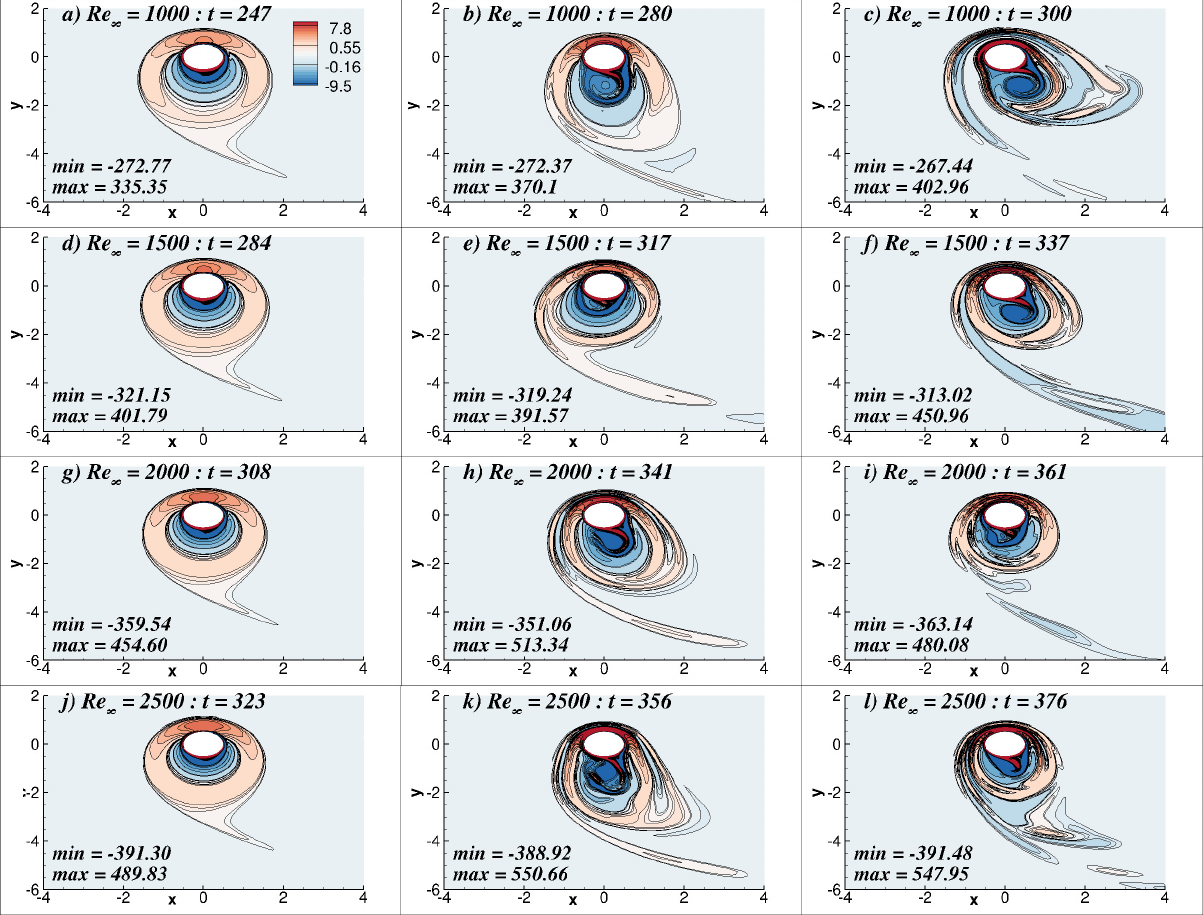}
%% Use \caption command for figure caption and label.
\caption{Time variation of spanwise vorticity for $U^*_s = 10U_{\infty}$ and (a)-(c)$Re_{\infty} = 1000$, (d)-(f)$Re_{\infty} = 1500$, (g)-(i) $Re_{\infty} = 2000$, (j)-(l) $Re_{\infty} = 2500$. The vorticity is plotted at indicated times starting from the onset of instability in frames (a), (d), (g), and (j).}
\label{fig3}
\end{figure*}

\begin{figure*}%% placement specifier
\centering
\includegraphics[width=\textwidth]{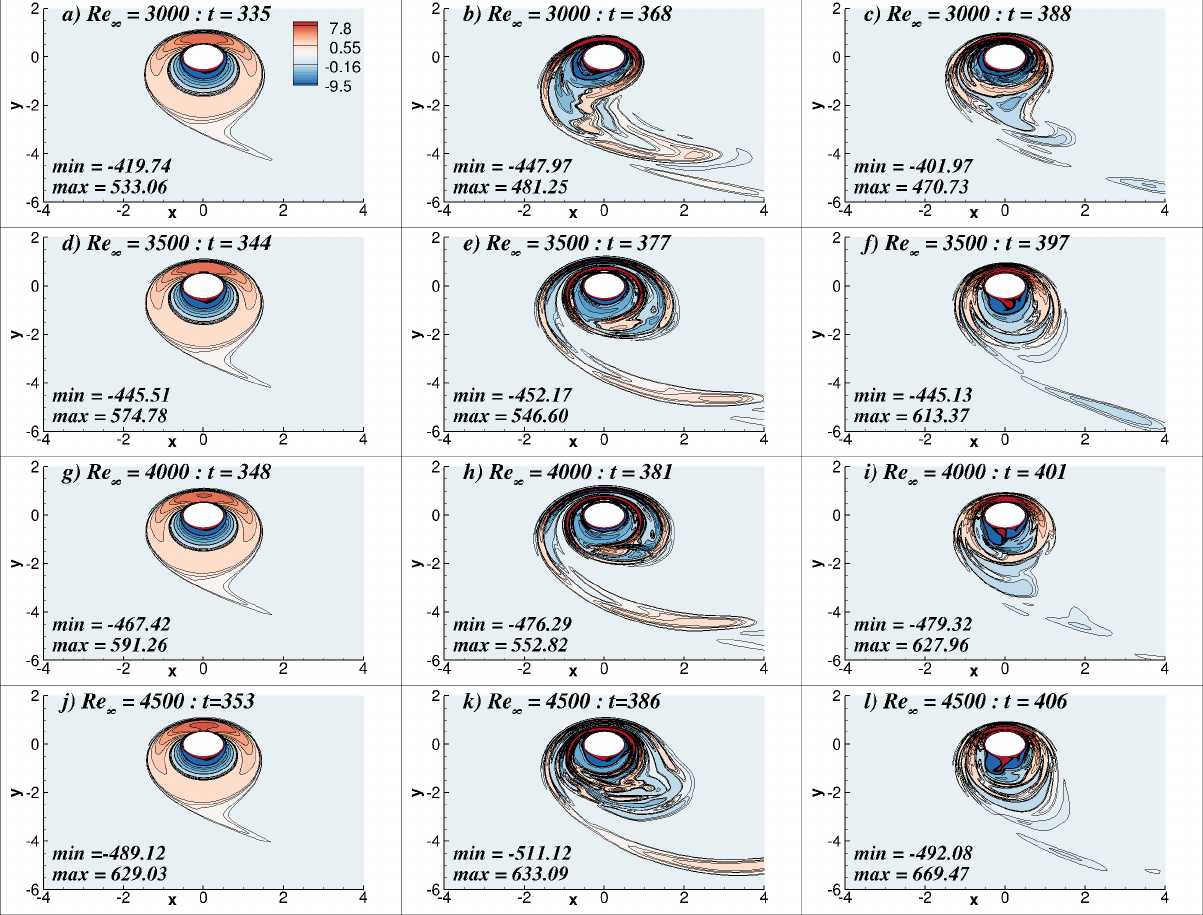}
%% Use \caption command for figure caption and label.
\caption{Time variation of spanwise vorticity for $U^*_s = 10U_{\infty}$ and (a)-(c)$Re_{\infty} = 3000$, (d)-(f)$Re_{\infty} = 3500$, (g)-(i) $Re_{\infty} = 4000$, (j)-(l) $Re_{\infty} = 4500$. The vorticity is plotted at indicated times starting from the onset of instability in frames (a), (d), (g), and (j).}
\label{fig4}
\end{figure*}

\begin{figure*}%% placement specifier
\centering
\includegraphics[width=\textwidth]{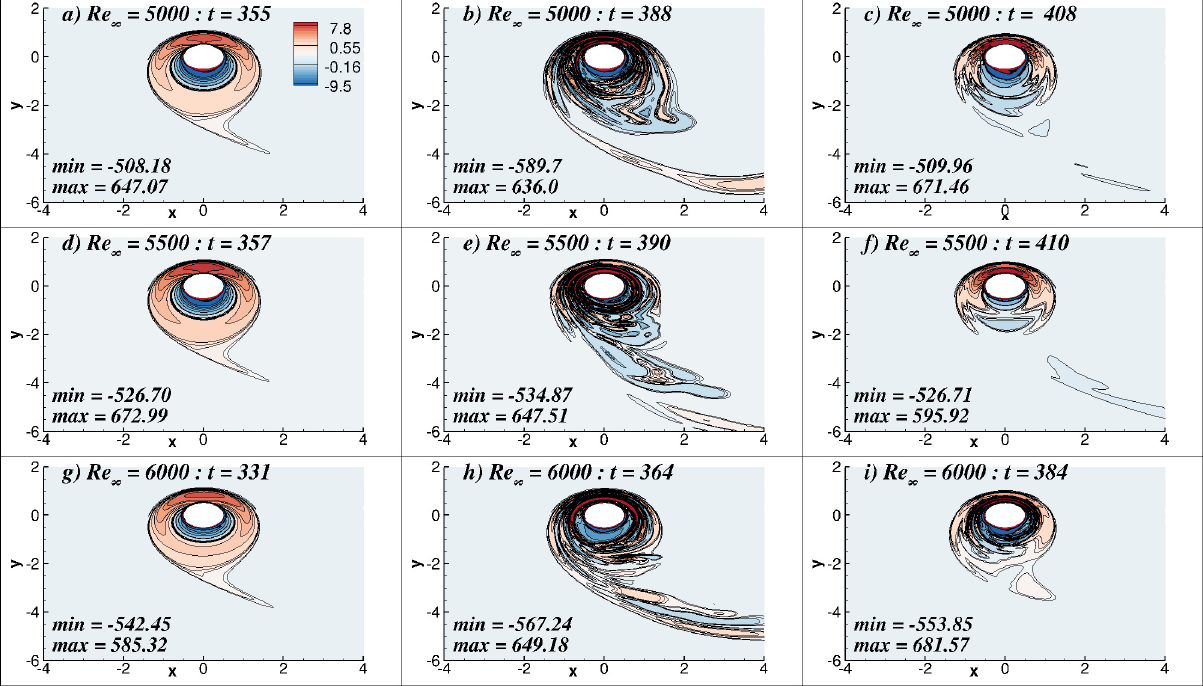}
%% Use \caption command for figure caption and label.
\caption{Time variation of spanwise vorticity for $U^*_s = 10U_{\infty}$ and (a)-(c)$Re_{\infty} = 5000$, (d)-(f)$Re_{\infty} = 5500$, (g)-(i) $Re_{\infty} = 6000$. The vorticity is plotted at indicated times starting from the onset of instability in frames (a), (d), and (g).}
\label{fig5}
\end{figure*}

The contour plots of spanwise vorticity are shown in Figs. \ref{fig3} to \ref{fig5} for $Re_{\infty}$ in the range of 1000 to 6000 in intervals of 500. The first frame for all $Re_{\infty}$ depicts the time at which the temporal instability commences (captured from the time-series). The two subsequent frames are chosen to provide a qualitative comparison of different $Re_{\infty}$ and the role it has on the ensuing vortical field in the proximity of the rotating cylinder.

In Fig. \ref{fig3}, the contours of spanwise vorticity are shown for $Re_{\infty} = 1000$, 1500, 2000, and 2500 in frames (a-c), (d-f), (g-i), (j-l), respectively. At the onset of the temporal instability, the vorticity field shows identical features for all $Re_{\infty}$ considered. The magnitude, however, increases with an increase in $Re_{\infty}$. At onset, direction of rotation of the cylinder, along with incoming flow convecting from left to right causes maximum positive vorticity on top part of the cylinder. The negative vorticity, on the other hand, is on the bottom half of the cylinder. This equilibrium state is not sustained, however, and positive vorticity accumulates and grows in the lower part of the flow (marked by dark red contours) near the cylinder surface, as seen in Figs. \ref{fig3}(b), \ref{fig3}(e), \ref{fig3}(h), and \ref{fig3}(k). This in turn, pushes out the patch of negative vorticity and causes it to eject from the bottom surface, which is seen in Figs. \ref{fig3}(c) and \ref{fig3}(f). A wedge appears in the negative vorticity region in the near-field of the cylinder. The trapped negative vorticity is ejected from the recirculating region, gyrating from the rotating cylinder \citep{sundaram2021multiscale}. For the higher $Re_{\infty}$ of 2000 and 2500 in Figs. \ref{fig3}(i) and \ref{fig3}(l) , the ejection of negative vorticity has already taken place by $t = 361$ and 376, respectively and the system has reached another quasi-steady equilibrium state. 

In Fig. \ref{fig4}, the contours of spanwise vorticity are shown for $Re_{\infty} = 3000$, 3500, 4000, and 4500 in frames (a-c), (d-f), (g-i), (j-l), respectively. As in Fig. \ref{fig3}, the onset of the temporal instability shows the typical asymmetry in strength and location of positive and negative vortices for flow past a rotating cylinder. The vorticity generated on the windward side of the spinning cylinder protrudes from the surface of the cylinder, named in the literature as a \lq tongue'-like structure \citep{mittal2003flow}. This structure protrudes upward and the length increases with $Re_{\infty}$, as seen in Figs. \ref{fig4}(b), \ref{fig4}(e), \ref{fig4}(h), and \ref{fig4}(k). As the $Re_{\infty}$ is increased progressively, the negative and positive vorticity patches are wrapped around the cylinder as tightly wound spirals. For the lower $Re_{\infty}$ in Fig. \ref{fig3}, these opposite signed vorticity appeared as \lq blobs' loosely surrounding the cylinder. The system evolves through multiple quasi-steady equilibrium states. All $Re_{\infty}$ are at one such equilibrium state in Figs. \ref{fig4}(c), \ref{fig4}(f), \ref{fig4}(i), and \ref{fig4}(l), wherein positive vorticity is once again accumulating on the bottom surface of the cylinder.

In Fig. \ref{fig5}, the contours of spanwise vorticity are shown for $Re_{\infty} = 5000$, 5500, and 6000 in frames (a-c), (d-f), and (g-i), respectively. The vortical flow field is distinctly different for these higher $Re_{\infty}$ compared to those in Figs. \ref{fig3} and \ref{fig4}. Apart from the onset of the temporal instability, subsequent vorticity dynamics shows an absence of the negative and positive vorticity regions in the near neighborhood of the cylinder surface. Regions of negative and positive vorticity are tightly wound on the cylinder surface, as noted in Figs. \ref{fig5}(b), \ref{fig5}(e), and \ref{fig5}(h). Fine-scale low-magnitude vortical structures eject from the bottom half of the cylinder, which do not retain their coherence in the unsteady flow field for long. This indicates that as $Re_{\infty}$ increases, unsteadiness increases in the flow field with multiple vortical structures in the wake of the spinning structure. It is thus, expected that vorticity will be distributed over multiple spatial and temporal scales as $Re_{\infty}$ increases. Frequency peaks associated with shed vortices will lose their prominence with increase in $Re_{\infty}$. At later stage in Figs. \ref{fig5}(c), \ref{fig5}(f), and \ref{fig5}(i), the system reaches a quiescent state with surface vorticity mostly accumulated on the top of the cylinder. The peak value of positive vorticity, for all $Re_{\infty}$, is roughly at the same location on the lower surface of the cylinder. It is the peak value of negative vorticity which varies significantly with $Re_{\infty}$. For low $Re_{\infty}$ (< 3500), this peak is observed on the windward side. As $Re_{\infty}$ increases, the new peak location starts to form at a location on the leeward side. This redistribution of negative vorticity at surface is perhaps what provides higher stability at higher $Re_{\infty}$. 

\subsection{Temporal scale selection}

The vorticity distribution shows the presence of multiple vortical eddies detaching from the bottom half of the cylinder. In this section, we quantify the temporal scales in the flow field as a function of $Re_{\infty}$, capturing the dominant peaks in the frequency plane and associated Fourier amplitudes. The sampling point for the time-series of vorticity is $x = 0, y = -1$, near the bottom surface of the spinning cylinder. As the rotating cylinder modifies the base flow through Magnus-Robins effect, it is observed that the upper side (where the surface moves with the free stream) has higher velocity and lower pressure, whereas the lower side (where the surface moves against the free stream) has lower velocity and higher pressure. This makes the bottom region dynamically complex due to separation of separated shear layer and strong vortex formation. It has been observed that fluctuations in lift and drag will originate downstream of the lower separation point \cite{suman2022novel}. Hence, sampling near the bottom surface of the rotating cylinder helps capture unsteady pressure (or velocity) signatures that are most representative of the flow instabilities caused by rotation.

\begin{figure*}%% placement specifier
\centering
\includegraphics[width=.9\textwidth]{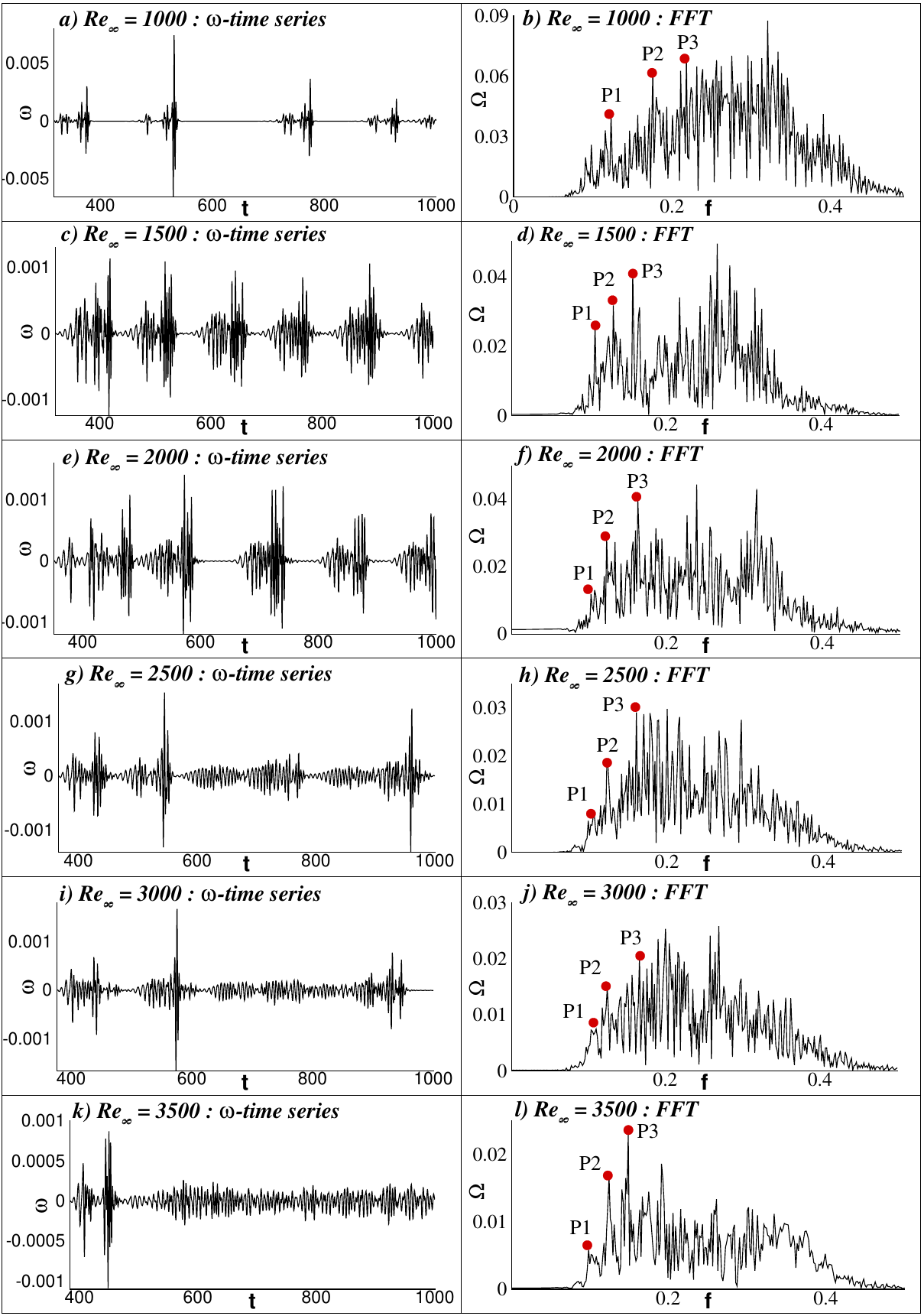}
%% Use \caption command for figure caption and label.
\caption{Time-series of spanwise vorticity probed at $x = 0, y = -1$ for (a) $Re_{\infty} = 1000$, (c) $Re_{\infty} = 1500$, (e) $Re_{\infty} = 2000$, (g) $Re_{\infty} = 2500$, (i) $Re_{\infty} = 3000$, and (k) $Re_{\infty} = 3500$, truncated to only show the growth of disturbance. The corresponding spectra are shown in frames (b), (d), (f), (h), (j), and (l), respectively.}
\label{fig6}
\end{figure*}

In Fig. \ref{fig6}, the time-series of spanwise vorticity is probed for $Re_{\infty} = 1000$, 1500, 2000, 2500, 3000, and 3500. The corresponding FFT of the time-series are shown on the right-side frames. For all the $Re_{\infty}$ shown here, flow reaches a temporally quasi-periodic solution, as indicated by the presence of multiple independent frequencies in the spectra. There is no case where a steady state is reached, as the dimensionless rotation rate is high. For flow with $Re_{\infty} = 1000$ in Figs. \ref{fig6}(a)-(b), the magnitude of the vorticity and its Fourier amplitude is higher than all other $Re_{\infty}$. As seen in Figs. \ref{fig3} to \ref{fig5}, as the $Re_{\infty}$ is increased, vorticity is redistributed along the bottom half of the rotating cylinder. This leads to weaker localised vorticity magnitude and lower amplitudes of vorticity in the spectral plane. The dominant peaks in the frequency plane are identified as P1, P2, P3 and these are recorded in Table \ref{tab1}. As $Re_{\infty}$ is increased to 1500, the peaks identified are of smaller amplitude, which shows that vorticity is being distributed to higher frequency. The contour plots in Fig. \ref{fig3} also showed that with increase in $Re_{\infty}$, there is an increased rightward deflection of vortical structures in the wake of the spinning cylinder. For example, Figs. \ref{fig3}(b), \ref{fig3}(e), \ref{fig3}(h), and \ref{fig3}(k), reveal a gradual lateral spread of the wake structures with an increase in $Re_{\infty}$. The maximum deflection increased to $x = 2.5$ for $Re_{\infty} = 2500$ from $x = 1.9$ for $Re_{\infty} = 1900$. This increase in lateral width of the wake with more number of fine-scale vortical ejections for higher $Re_{\infty}$, leads to a redistribution of the vorticity. Thus, the intensity of the vortices shed is reduced and more temporal scales are introduced in the flow. Similar observations can be made for $Re_{\infty}$ in the range of 2000-3500. The secondary peaks, P2 and P3 also become sub-dominant as $Re_{\infty}$ increases, with a broader range of higher frequencies where the amplitude is higher. This can be explained by patches of negative vorticity being ejected from surface of the cylinder in multiple layers, as $Re_{\infty}$ increases. Thus, formation of multiple vortical eddies explains the multimodal \cite{sengupta2020effects} nature of the spectrum.

\begin{table}
\begin{center}
\def~{\hphantom{0}}
\caption{Dominant frequencies and their amplitudes, marked as P1, P2, P3 in Figs. \ref{fig6} and \ref{fig7}}
\begin{tabular}{lccccccc}
 \hline
$Re_{\infty}$ & P1($f$) & P1($\Omega$) & P2($f$) & P2($\Omega$) & P3($f$) & P3($\Omega$)  \\ [3pt] 
 \hline 
 1000 & 0.12298 & 0.04071 & 0.17569 & 0.06144 & 0.21815 & 0.06985 \\ 
 \hline
 1500 & 0.10882 & 0.02608 & 0.13235 & 0.03164 & 0.157353 & 0.04137 \\
 \hline
 2000 & 0.10785 & 0.01303 & 0.12326 & 0.02884 & 0.16332 & 0.03976 \\
 \hline
 2500 & 0.99368 & 0.00725 & 0.12303 & 0.01804 & 0.16088 & 0.02915 \\
 \hline
 3000 & 0.10417 & 0.00685 & 0.12500 & 0.01481 & 0.16666 & 0.02028 \\
 \hline
 3500 & 0.09902 & 0.00538 & 0.12500 & 0.01667 & 0.14935 & 0.02293 \\
 \hline
 4000 & 0.10098 & 0.00529 & 0.12578 & 0.01384 & 0.14728 & 0.02540 \\
 \hline
 4500 & 0.099831 & 0.00520 & 0.11980 & 0.00839 & 0.14808 & 0.01746 \\
 \hline
 5000 & 0.10099 & 0.00436 & 0.12913 & 0.01196 & 0.15397 & 0.01772 \\
 \hline
 5500 & 0.10429 & 0.00339 & 0.12914 & 0.00782 & 0.14403 & 0.01461 \\
 \hline
 6000 & 0.11624 & 0.00375 & 0.12420 & 0.00636 & 0.14649 & 0.02029 \\
 \hline
\end{tabular}
\label{tab1}
\end{center}
\end{table}

\begin{figure*}%% placement specifier
\centering
\includegraphics[width=.9\textwidth]{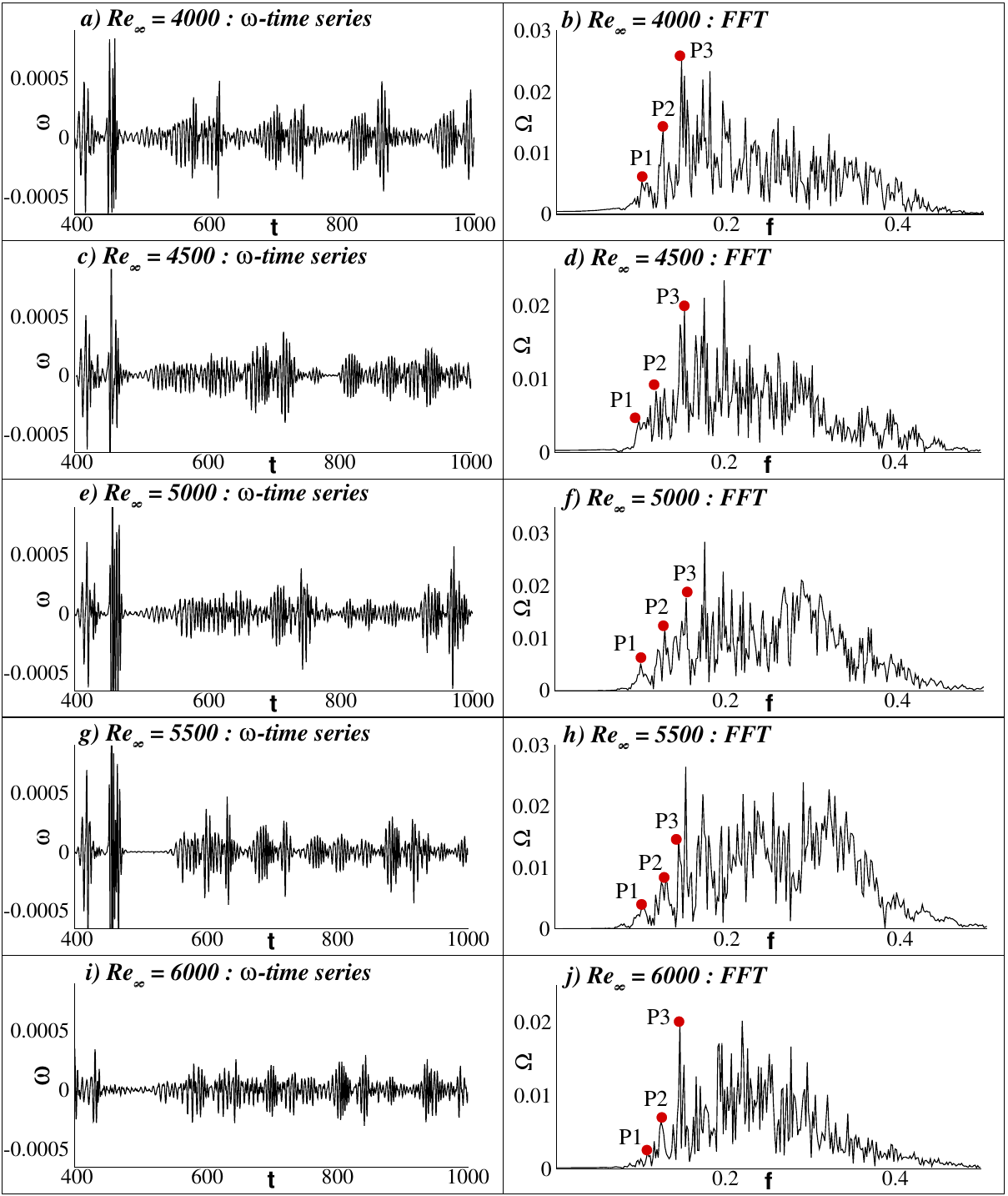}
%% Use \caption command for figure caption and label.
\caption{Time-series of spanwise vorticity probed at $x = 0, y = -1$ for (a) $Re_{\infty} = 4000$, (c) $Re_{\infty} = 4500$, (e) $Re_{\infty} = 5000$, (g) $Re_{\infty} = 5500$, and (i) $Re_{\infty} = 6000$, truncated to only show the growth of disturbance. The corresponding spectra are shown in frames (b), (d), (f), (h), (j), and (l), respectively.}
\label{fig7}
\end{figure*}

In Fig. \ref{fig7}, the time-series and FFT of spanwise vorticity, probed for $Re_{\infty} = 4000$, 4500, 5000, 5500, and 6000. Upon further increase of $Re_{\infty}$ to 4000, the vorticity magnitude and its Fourier amplitude decreases further with numerous peaks in the high frequency range. These peaks have equivalent amplitude to the three major peaks, P1-P3 for $Re_{\infty} > 5000$. As seen in Figs. \ref{fig4} to \ref{fig5}, as $Re_{\infty}$ is increased, vorticity is redistributed along the bottom half of the rotating cylinder, and the time period of oscillations decreases as the oblique instability wave appears. The frequency of vortex shedding also increases, which leads to high amplitude peaks, as seen in Fig. \ref{fig7}(b). The dominant peaks in the frequency plane are identified as P1, P2, and P3 and are provided in Table \ref{tab1}. At higher $Re_{\infty}$, there is  a longer time period of vortex shedding due to nonlinear interactions between shear layers. The contour plots in Fig.\ref{fig4} also show that there is a rightward deflection of the vortical structures in the wake of the spinning cylinder, as $Re_{\infty}$ increases beyond 4500. For flow with $Re_{\infty}$ in the range of 5000-6000, with an increase in Reynolds number, the compressibility effect dominates and inhibits the mixing between shear layers. This leads to delayed vortex shedding with most dominant peak noted at higher frequencies (with smaller time period of shedding). The contour plots in Fig.\ref{fig5}, also showed that the vortical structures in the wake are redistributed towards the bottom of the cylinder with a decrease in width. For $Re_{\infty} = 6000$, there is a suppression of vortex shedding as the wake behind the cylinder transitions from a periodic oscillatory state to aperiodic oscillatory state with more symmetric flow. This observation is consistent with the contour plots shown in Fig.\ref{fig5}.

Vorticity fluctuations (and flow unsteadiness) are the strongest for lower $Re_{\infty}$ due to the interplay of viscous diffusion, rotation, and wake asymmetry. At lower $Re_{\infty}$, viscous forces are significant, but rotation strongly distorts the near-wall shear layers by introducing opposite-signed vorticity layers on the upper and lower sides of the cylinder. Viscosity is not strong enough to diffuse the vortical layers before they interact downstream. Instead, rotation causes a strong shear and roll-up of vorticity near the separation region, producing large coherent vortices that fluctuate periodically \cite{chew1995numerical}. While the overall kinetic energy is small, the relative intensity of vorticity fluctuations is large, for lower $Re_{\infty}$. In contrast, as $Re_{\infty}$ increases, relative influence of rotation decreases compared to inertial forces. Magnus-Robins effect becomes more stabilizing — rotation delays or even suppresses vortex shedding altogether (for sufficiently high rotation rates) \cite{karabelas2012high}. The wake becomes more asymmetric but steadier, with less periodic shedding and smaller coherent vortices. So, even though higher $Re_{\infty}$ flows have stronger inertial energy, rotation tends to stabilize them, leading to smaller vorticity fluctuations.

\subsection{Unsteady force distribution}

In this section, we evaluate the coefficients of lift and drag at various $Re_{\infty}$ for flow past a rotating cylinder to quantify and understand how rotation affects the aerodynamic forces acting on the body. A rotating cylinder creates asymmetric circulation around the body which results in a pressure difference on the upper and lower surfaces, producing lift. From the solution of Eq. \eqref{eq1}, the normal and shear stresses acting on the cylinder surface can be evaluated. Using this, the lift coefficient ($C_l$) which directly measures the strength of the Magnus-Robins effect \citep{white1994fluid}, is calculated as follows, $C_l = \frac{F_y}{0.5 \rho_{\infty} U_{\infty}^2 D}$. Similarly, the drag coefficient ($C_d$) is calculated as $C_d =  \frac{F_x}{0.5 \rho_{\infty} U_{\infty}^2 D}$, where $F_x$ and $F_y$ are the $x$- and $y$-components of the net force acting along the circumference of the cylinder. By studying $C_l$ and $C_d$ over a range of $Re_{\infty}$, we can understand the transition from symmetric to asymmetric wake, the onset of vortex shedding suppression and the conditions leading to lift and drag variation.

\begin{figure*}%% placement specifier
\centering
\includegraphics[width=.9\textwidth]{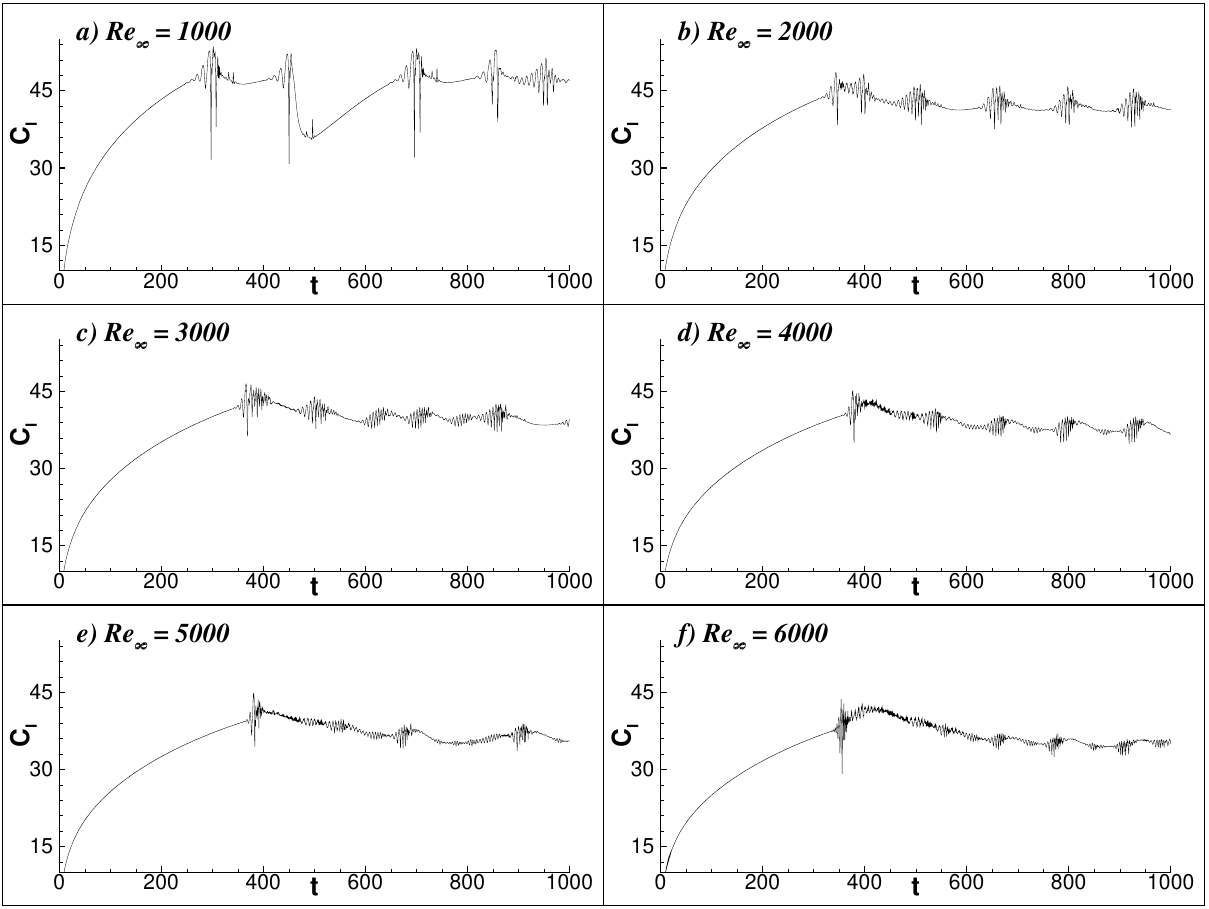}
%% Use \caption command for figure caption and label.
\caption{Time variation of lift coefficient for (a) $Re_{\infty} = 1000$, (b) $Re_{\infty} = 2000$, (c) $Re_{\infty} = 3000$, (d) $Re_{\infty} = 4000$, (e) $Re_{\infty} = 5000$, and (f) $Re_{\infty} = 6000$. }
\label{fig8}
\end{figure*}

In Fig. \ref{fig8}, the time variation of $C_l$ for a rotating cylinder with dimensionless rotation rate of $U_s^* = 10 U_{\infty}$ is compared for $Re_{\infty}$ = 1000 to 6000, in intervals of 1000. Here, it can be seen that $C_l$ appears to plateau or decrease with an increase in $Re_{\infty}$. This can be interpreted as follows: at high $Re_{\infty}$, inertial forces dominate, such that even if the rotation rate is high, the relative influence of rotation-induced circulation becomes weaker compared to the increasing free-stream inertia. The $C_l$, which is normalized by dynamic pressure, decreases because the freestream velocity $U_{\infty}$ is increasing with $Re_{\infty}$, but circulation may not increase proportionally. The circulation generated by rotation does not increase linearly with $Re_{\infty}$ and for higher $Re_{\infty}$, vortex shedding may be suppressed or altered, and circulation may saturate, even if the rotation rate is constant \citep{tokumaru1991rotary}. The time variation of $C_l$ in Figs. \ref{fig8}(d)-(f) is much less chaotic with fewer time-scales compared to the lower $Re_{\infty}$. At high $Re_{\infty}$, such as those shown in Fig. \ref{fig5}, especially with high rotation rate, the flow becomes more stable and separation is suppressed. While this occurrence stabilizes the flow and reduces drag, it can also reduce the asymmetry in the pressure field, which is needed for high lift. Furthermore, at high $Re_{\infty}$, the top shear layer separates laminarly from the cylinder and transitions to turbulence downwind. This transition has been reported in a direct numerical simulation by Aljure {\it et al.} \cite{aljure2015influence}, where flow past a rotating cylinder was simulated for $Re_{\infty} = 5000$ and dimensionless rotation rates 1 and 2. The dominant mechanism for transition was attributed to Kelvin-Helmholtz instabilities with growing \lq crescent-shaped' vortical structures. Furthermore, the bottom shear layer transitions to turbulence with point of transition in windward side of the cylinder. The presence of fully turbulent wake reduces the difference in pressure between the top and bottom surfaces, as previously reported by Stojković {\it et al.} \cite{stojkovic2002effect}. This leads to a milder Magnus-Robins effect and thus reduced lift generation.

\begin{figure*}%% placement specifier
\centering
\includegraphics[width=.9\textwidth]{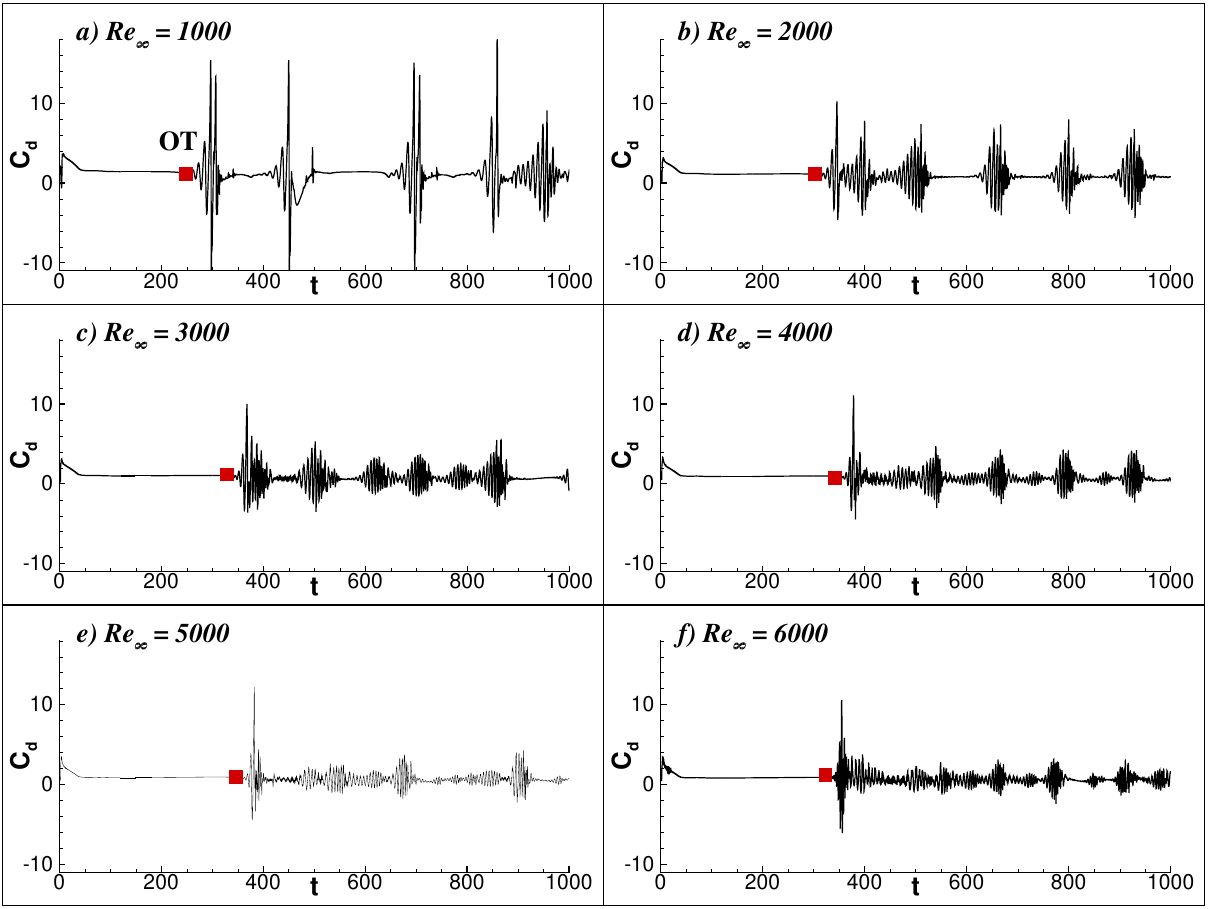}
%% Use \caption command for figure caption and label.
\caption{Time variation of drag coefficient for (a) $Re_{\infty} = 1000$, (b) $Re_{\infty} = 2000$, (c) $Re_{\infty} = 3000$, (d) $Re_{\infty} = 4000$, (e) $Re_{\infty} = 5000$, and (f) $Re_{\infty} = 6000$. The onset time of instability is marked as OT in the figure.}
\label{fig9}
\end{figure*}

In Fig. \ref{fig9}, the time variation of $C_d$ for a rotating cylinder with high dimensionless rotation rate of $U_s^* = 10 U_{\infty}$ is compared for $Re_{\infty}$ = 1000 to 6000, in intervals of 1000. Here, we also record the onset time of the instability due to Magnus-Robins effect by OT in the figure. It is evident that the drag coefficient reduces in amplitude with an increase in $Re_{\infty}$. As $Re_{\infty}$ increases, the boundary layer on the cylinder surface transitions from laminar to turbulent, which has more momentum near the wall and thus is resistant to separation. This delay in flow separation reduces the wake size and pressure drag, which is the dominant drag component at moderate to high $Re_{\infty}$. This is much more pronounced for the higher $Re_{\infty}$ shown in Fig. \ref{fig5}. For very high rotation rates, there is a strong tangential velocity induced on the cylinder surface, which promotes a favorable pressure gradient on the advancing side of the cylinder. Thus, the flow remains attached for longer at higher $Re_{\infty}$ (where boundary layers are thinner and more receptive to rotation effects). The combined effect of rotation and high $Re_{\infty}$ contributes to suppressing flow separation, leading to a narrower wake, reduced vortex shedding and a lower pressure differential. The cleaner and more symmetric wake results in lower drag formation. It is expected that the viscous drag will increase with $Re_{\infty}$, but the pressure drag dominates for flow past a rotating cylinder. Since pressure drag drops sharply with the shrinkage of the wake structures, the net $C_d$ drops despite a slight rise in friction.

At lower $Re_{\infty}$, vortex shedding takes place from the top and bottom of the stationary cylinder which creates periodic oscillations in lift and drag via the K\'{a}rm\'{a}n vortex street. With rotation introduced across the cylinder surface, this shedding is weakened or eliminated. When $Re_{\infty}$ increases for the rotating cylinder, the boundary layer becomes turbulent and more stable, and combined with high rotation, vortex shedding is suppressed altogether. For high rotation rates and high $Re_{\infty}$, we observed a narrow, symmetric, and attached wake, as seen in Figs. \ref{fig5}(c), \ref{fig5}(f), and \ref{fig5}(i). Such wake structures produce minimal temporal disturbances leading to a reduction in the unsteadiness with the system becoming \lq quasi'-steady or even fully steady. The small-scale instabilities in the shear layers or separated regions are damped due to increased inertia and higher momentum in the boundary layer. Thus, we recover smoother time histories for $C_l$ and $C_d$ in Figs. \ref{fig8} and \ref{fig9} at higher $Re_{\infty}$.

The time-averaged $C_l$ and $C_d$ are computed for the $Re_{\infty}$ reported in Figs. \ref{fig8} and \ref{fig9}, and are tabulated in Table \ref{tab2}. The trend of decreasing $\overline{C_l}$ and $\overline{C_d}$ with increasing $Re_{\infty}$ is explained while conducting the bifurcation analysis, in the next section.

\begin{table}
\begin{center}
\def~{\hphantom{0}}
\caption{Time-averaged $C_l$ and $C_d$ for the various test cases shown in Figs. \ref{fig8} and \ref{fig9}}
\begin{tabular}{lccc}
 \hline
$Re_{\infty}$ & $\overline{C_l}$ & $\overline{C_d}$ \\ [3pt] 
 \hline 
 1000 & 42.8792 & 1.1969 \\ 
 \hline
 2000 & 39.5816 & 0.9743 \\
 \hline
 3000 & 37.2986 & 0.8838  \\
 \hline
 4000 & 35.8885 & 0.8183  \\
 \hline
 5000 & 34.5318 & 0.7578  \\
 \hline
 6000 & 33.8978 & 0.7337  \\
 \hline
 \end{tabular}
\label{tab2}
\end{center}
\end{table}

\subsection{Bifurcation analysis at high dimensionless rotation rate}

In this section, the variation of lift, drag, and the onset time for the instability (as marked in Fig. \ref{fig9}) with $Re_{\infty}$ is explored for a fixed high dimensionless rotation rate. This serves as a benchmark dataset for testing the efficacy of compressible Navier-Stokes equation solvers, validating turbulence models, and for studying flow separation and reattachment. The maximum or root mean square (RMS) values of $C_l$ and $C_d$ reflect the bifurcation amplitude, which in turn defines the bifurcation point. Plotting maximum $C_l$ or $C_d$ versus $Re_{\infty}$ therefore makes the bifurcation curve (branch) clearly visible. In some studies \cite{cheng2007characteristics} RMS or Fourier amplitude of $C_l$ and $C_d$ are used, which represent the same physical idea — the magnitude of unsteady oscillations. However, using maximum (peak value) is simpler to extract from transient or noisy data, easier to identify critical onset in experiments, and yet still proportional to the oscillation amplitude for small periodic signals.

\begin{figure*}%% placement specifier
\centering
\includegraphics[width=.8\textwidth]{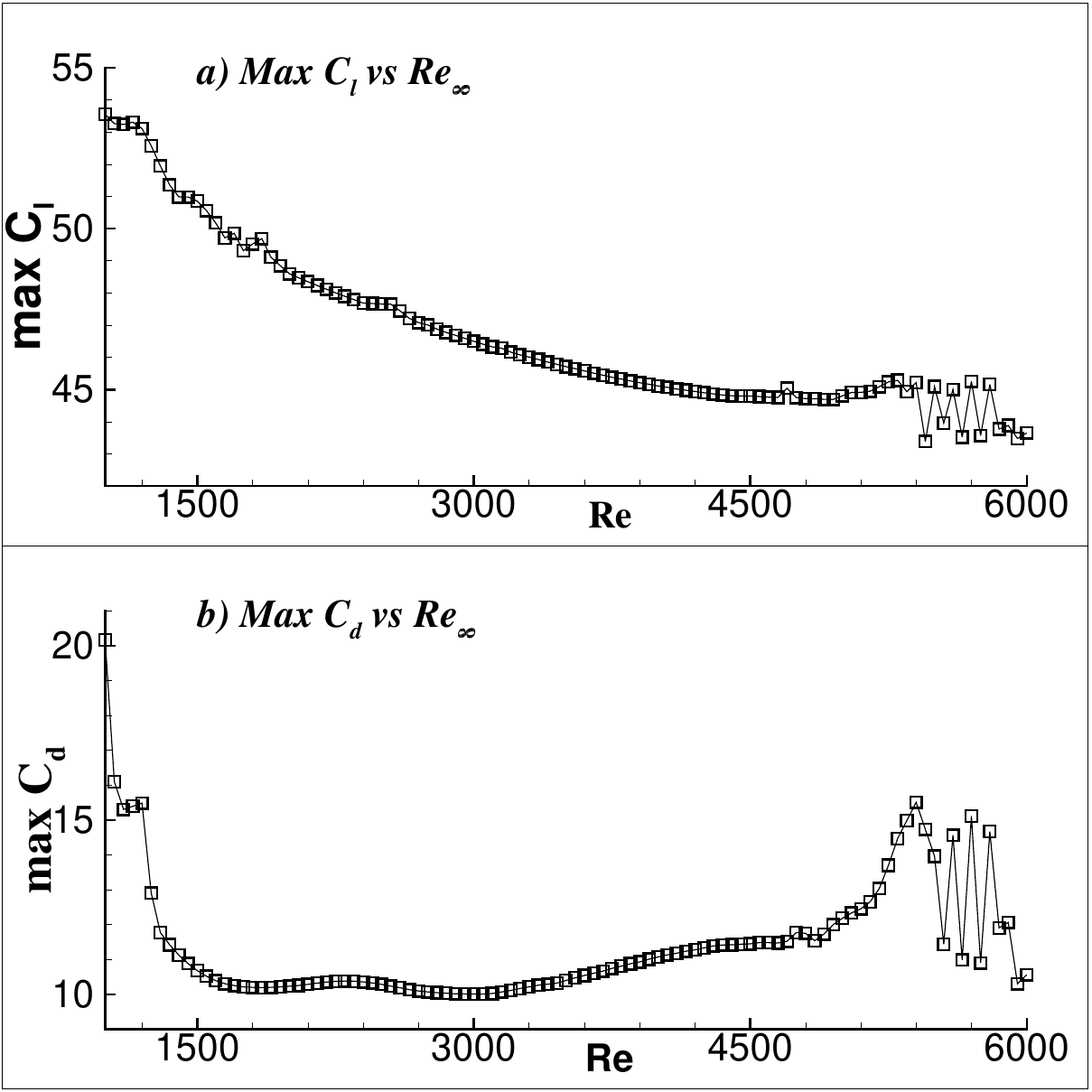}
%% Use \caption command for figure caption and label.
\caption{Variation of maximum coefficients of (a) lift and (b) drag, as a function of $Re_{\infty}$ for flow past rotating cylinder with dimensionless rotation rate, $U^*_s=10U_{\infty}$.}
\label{fig10}
\end{figure*}

In Fig. \ref{fig10}, the variation of the maximum $C_l$ and maximum $C_d$ with $Re_{\infty}$ are shown for a fixed dimensionless rotation rate. In Fig. \ref{fig10}(a), till $Re_{\infty} = 1250$, there is a flat plateau in the lift distribution. This can be explained by the fact that for $Re_{\infty} \le 1250$, the rotation is strong enough to generate a stable circulation, leading to high lift. Rotation dominates flow dynamics over the inertial force due to low $Re_{\infty}$. The boundary layer near the bottom surface of the spinning cylinder remains attached due to the rotation-induced favorable pressure gradient. It is not receptive to the temporal instability and the time variation reveals the presence of fewer time scales. Thus, the lift remains nearly constant as neither transition nor new instabilities have influenced the lift. However as $Re_{\infty}$ is increased further, the boundary layer becomes thinner and the shear layer is receptive to instabilities. Despite the high rotation rate, the circulation generation saturates. Localized disturbances lead to small-scale unsteadiness in the wake of the cylinder, which reduce the net pressure asymmetry around the cylinder. For the moderate $Re_{\infty}$, viscous effects compete with the stabilizing influence of rotation. Thus, there is a gradual loss in lift generation due to asymmetry in the pressure field, flow separation or vortex leakage near stagnation zones near the bottom surface of the cylinder, and primarily a reduced effectiveness of the Magnus-Robins effect for $1250 < Re_{\infty} < 5650$. Beyond $Re_{\infty} = 5650$, transition to unsteady or turbulent wake flow occurs. The flow is receptive to time-dependent instabilities and wake structures become dynamically unstable, such that even small changes in $Re_{\infty}$ lead to different instantaneous force distributions. Similar transitional wake dynamics and lift drop was observed \cite{mittal2003flow} near $Re_{\infty} \approx 5000$ for a much lower rotation rate. 

The $C_d$ variation in Fig. \ref{fig10}(b) shows a sharp drop till $Re_{\infty} = 1250$, as flow is strongly influenced by viscous forces for this relatively low value of $Re_{\infty}$. The rotation induces circulation that suppresses flow separation and reduction in vortical structures in the wake of the cylinder. This significantly reduces pressure drag, which dominates the drag distribution. Between $Re_{\infty} = 1250$ and 1500, there is a further drop in $C_d$ but this drop is more gradual. As inertial effects increase in prominence, the boundary layer becomes thinner, and flow separation is still suppressed, but the effect of rotation saturates. Drag continues to decrease, but more slowly, as viscous effects still decline with $Re_{\infty}$, but less dramatically. On further increase of $Re_{\infty}$ in the range $1500< Re_{\infty} \le 3500$, wake of the cylinder is a narrow and stable region, with vortex shedding being suppressed by rotation. However, the benefits of increasing $Re_{\infty}$ taper off, as separation has been suppressed to its maximum extent and the pressure recovery stalls. For $5650 > Re_{\infty} > 3500$, the flow becomes receptive to localized instabilities in the shear layer and in the wake of the cylinder. While rotation delays separation, small unsteady eddies or vortices form intermittently, on the advancing side of the cylinder. These disturbances increase mixing and momentum loss, causing earlier degradation in the pressure recovery. Beyond, a critical $Re_{\infty} = 5650$, flow enters a fully unsteady transitional regime. It exhibits bifurcation and chaotic wake states, highly sensitive to small changes in $Re_{\infty}$. Drag fluctuates due to changes in instantaneous pressure distribution, wake turbulence and bifurcation.

\begin{figure*}%% placement specifier
\centering
\includegraphics[width=.85\textwidth]{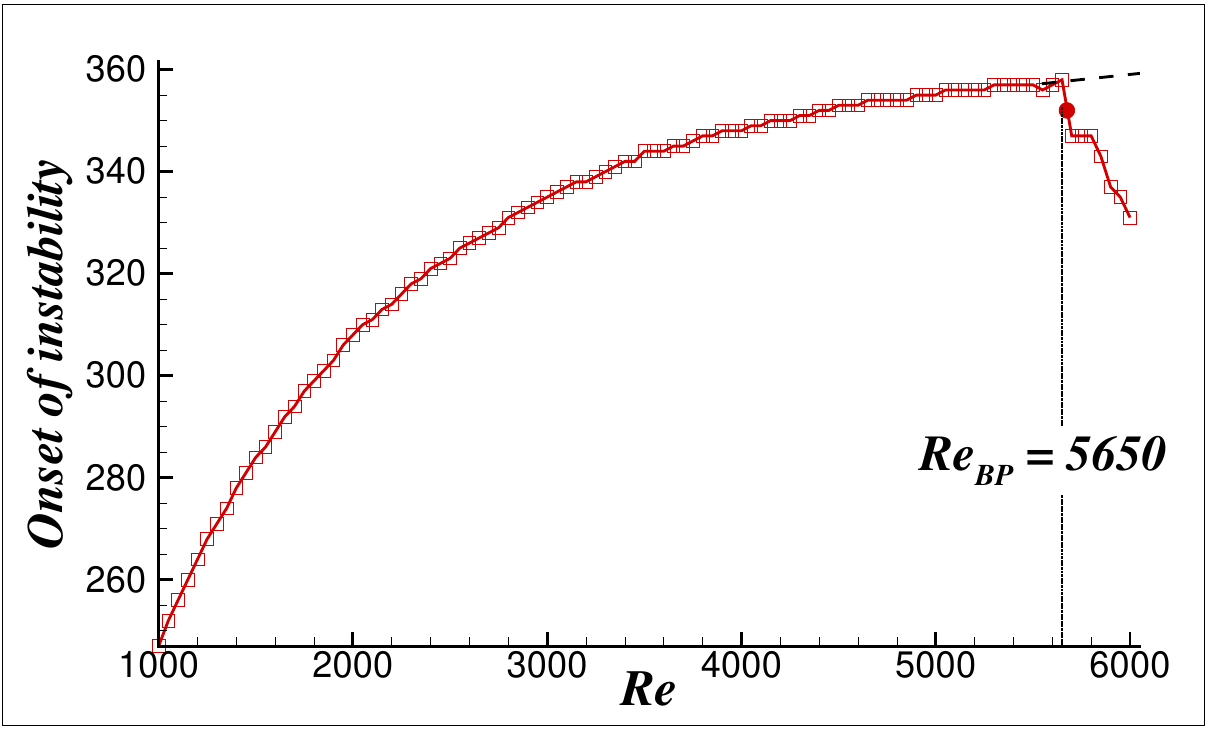}
%% Use \caption command for figure caption and label.
\caption{Variation of onset time of instability, as a function of $Re_{\infty}$ for flow past rotating cylinder with high dimensionless rotation rate, $U^*_s=10U_{\infty}$. The system bifurcates from the typical trend (shown by the dotted black line) beyond $Re_{BP} = 5650$.}
\label{fig11}
\end{figure*}

The onset time of the instability is captured from the time-series of $C_d$, as shown in Fig. \ref{fig9} as the time at which unsteady behavior (oscillations in drag) first appears in a previously steady flow. The variation of the onset time of instability with $Re_{\infty}$ is shown in Fig. \ref{fig11}. The onset time increases till the critical $Re_{\infty}$ value of 5650, which has been identified in the $C_l$ and $C_d$ variations in Fig. \ref{fig10}. Beyond this critical value, the system bifurcates and shows a sharp decline in the onset time. The dotted line follows the trajectory of the onset time variation prior to bifurcation. 

For a rotating cylinder exhibiting a high dimensionless rotation rate, a strong circulation is induced, leading to asymmetric pressure fields, suppressed separation and a stabilized wake near the bottom of the cylinder. This is consistent with a prior work \citep{suman2022novel} for flow past a rotating cylinder at high dimensionless rotation rates of $U_s^* = 12$ and 14. These delay the growth of disturbances in the shear layer, particularly at moderate Re. The flow resists unsteady behavior, so perturbations grow slowly, pushing instability onset to later times. As $Re_{\infty}$ increases, the boundary layer on the surface of the cylinder becomes thinner, shear layers destabilize and wake becomes unsteady with chaos eventually setting in. For subcritical $Re_{\infty}$, even when $Re_{\infty}$ is increased, the rotation rate is high enough that the side of the cylinder exhibiting co-rotation injects momentum into near wake. Rotational control becomes stronger as relative wall velocity introduces large tangential momentum compared to the inertia of the incoming flow. This suppresses the separation and vortex shedding is reduced. Thus, for $Re_{\infty} < 5650$, rotation still dominates the wake dynamics and the flow remains globally stable. When $Re_{\infty} = 5650$ is attained, flow structures in the wake of the cylinder and in the shear layer are receptive and small disturbances grow spatio-temporally till the system becomes globally unstable. The flow bifurcates into a new regime wherein the structures in the wake of the cylinder become chaotic. Further increment in $Re_{\infty}$ (and thus the inertial term) enhances inertial instabilities in the shear layers, promoting chaotic vortex shedding. This is a supercritical Hopf bifurcation \citep{theofilis2011global} wherein small changes in $Re_{\infty}$ cause large changes in flow behavior. Beyond $Re_{\infty} = 5650$, the growth rate of unstable modes increases sharply and the flow can no longer suppress disturbances with rotation alone. Global instability modes dominate the dynamics and the flow transitions from linearly stable to fully nonlinear and unsteady.

\subsubsection{Explaining the point of bifurcation}

Bifurcation analysis is critically important in flow past a rotating cylinder with a fixed high dimensionless rotation rate, as it reveals how a stable, steady flow can become unsteady, periodic, or chaotic as $Re_{\infty}$ changes. The previous discussion had led us to the critical $Re_{\infty}$ of 5650, beyond which periodic oscillations in the force distributions occurred. In this section, we will explore a sub-critical and super-critical $Re_{\infty}$ in addition to the critical value to explain how the instantaneous vorticity dynamics vary.

\begin{figure*}%% placement specifier
\centering
\includegraphics[width=.9\textwidth]{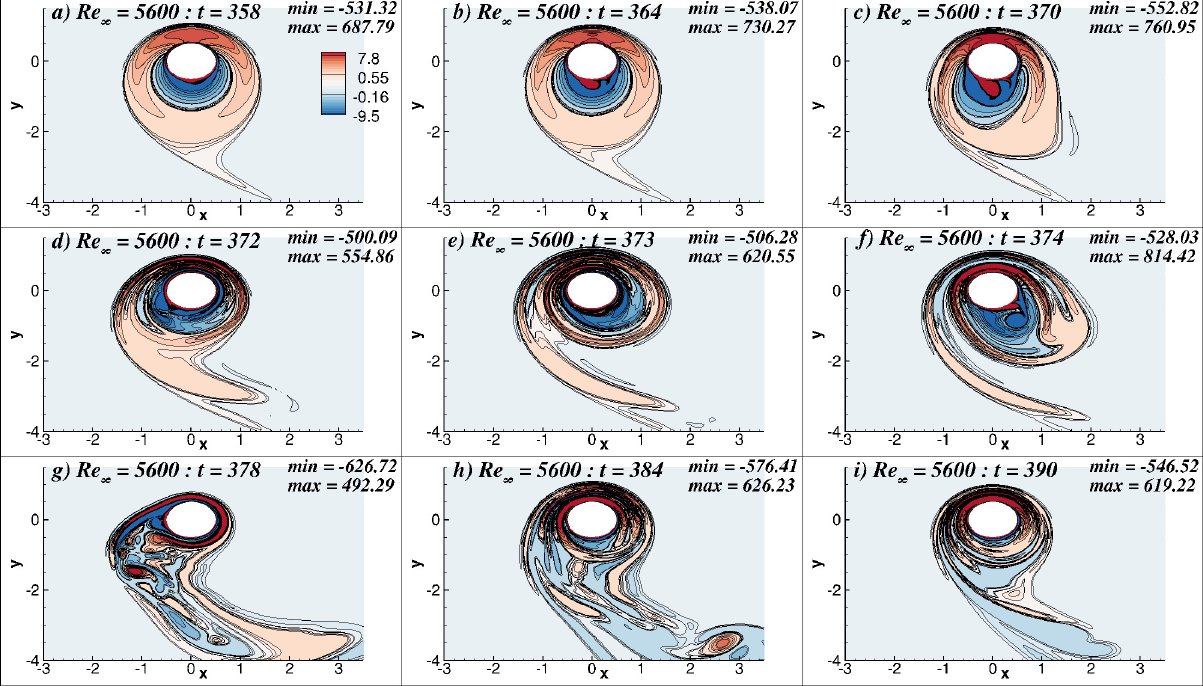}
%% Use \caption command for figure caption and label.
\caption{Spanwise vorticity contours for $U^*_s = 10U_{\infty}$ and sub-critical $Re_{\infty} = 5600$, shown at indicated times starting from the onset of instability.}
\label{fig12}
\end{figure*}

In Fig. \ref{fig12}, the contours of spanwise vorticity are shown at the indicated times for a sub-critical $Re_{\infty} = 5600$. The first frame is at the onset time of the instability. For a high rotation rate, the flow remains laminar and steady and the wake is not receptive to instability mechanisms. A strong asymmetric vorticity layer forms due to rotation with clockwise (positive) vorticity dominating on the advancing (top) side. Counter-clockwise (negative) vorticity accumulates near the retreating side. Here, vortical structures in the wake of the cylinder are attached and stabilized by rotation. The shear layer in the immediate vicinity of the bottom of the spinning cylinder is smooth and attached. There is a breakdown in symmetry due to rotation, but no time-dependent vortical structures are noted in the flow field. For instance, no Kelvin-Helmholtz roll-ups or vortex shedding is observed, only attached vorticity sheets are observed.

\begin{figure*}%% placement specifier
\centering
\includegraphics[width=.9\textwidth]{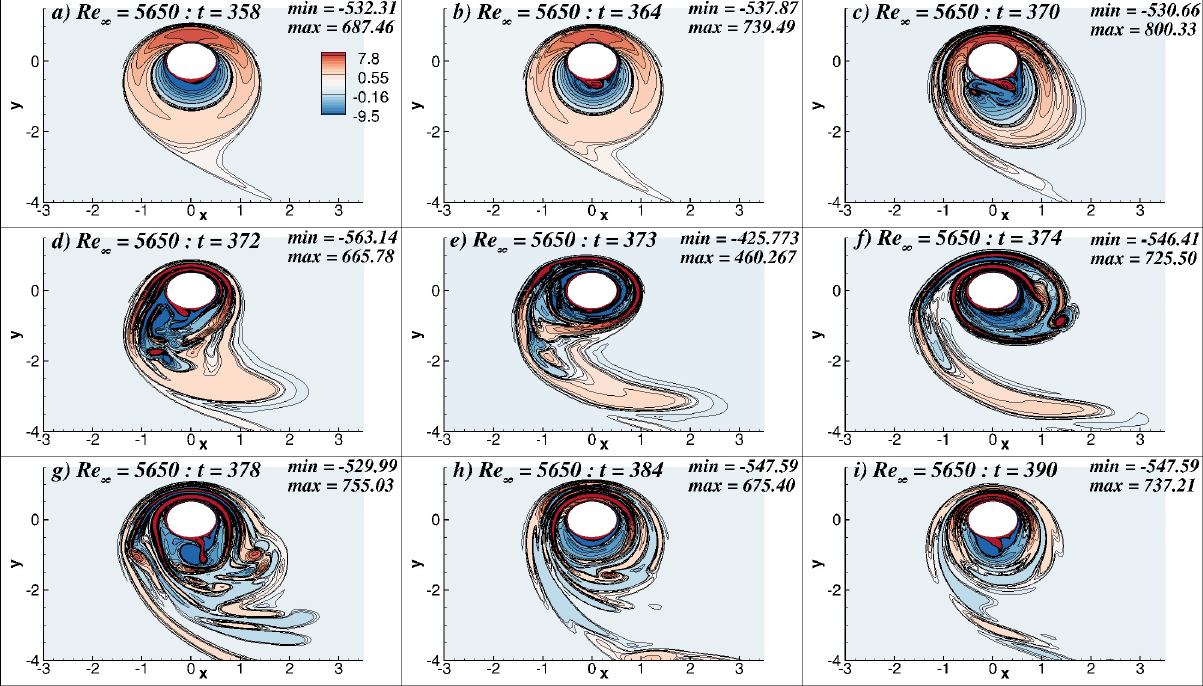}
%% Use \caption command for figure caption and label.
\caption{Spanwise vorticity contours for $U^*_s = 10U_{\infty}$ and critical $Re_{\infty} = 5650$, shown at indicated times starting from the onset of instability.}
\label{fig13}
\end{figure*}

In Fig. \ref{fig13}, the vorticity field is shown for the indicated times for a critical $Re_{\infty} = 5650$. The flow undergoes a Hopf bifurcation and the equilibrium state becomes unstable to oscillatory modes. Contrary to the sub-critical $Re_{\infty}$ plots shown in Fig. \ref{fig12}, the flow shows the presence of small, time-periodic disturbances near the top surface of the cylinder, as seen in Fig. \ref{fig13}(d). A global instability mode emerges in the shear layer and near wake, which leads to a sharpening of the \lq tongue' identified earlier in Figs. \ref{fig3} to \ref{fig5}. This is observed in Fig. \ref{fig13}(f), with a strong adherence of the vortical structures (initially small in amplitude) to the cylinder surface. Furthermore, weak unsteady vortical patches appear in the wake of the cylinder, as shown in Figs. \ref{fig13}(g) and \ref{fig13}(h). The shear layer has become sensitive to perturbations, with Kelvin–Helmholtz-type roll-ups initiated beyond $t = 374$. However, rotation is still dominant, so vortex shedding is modulated.

\begin{figure*}%% placement specifier
\centering
\includegraphics[width=.9\textwidth]{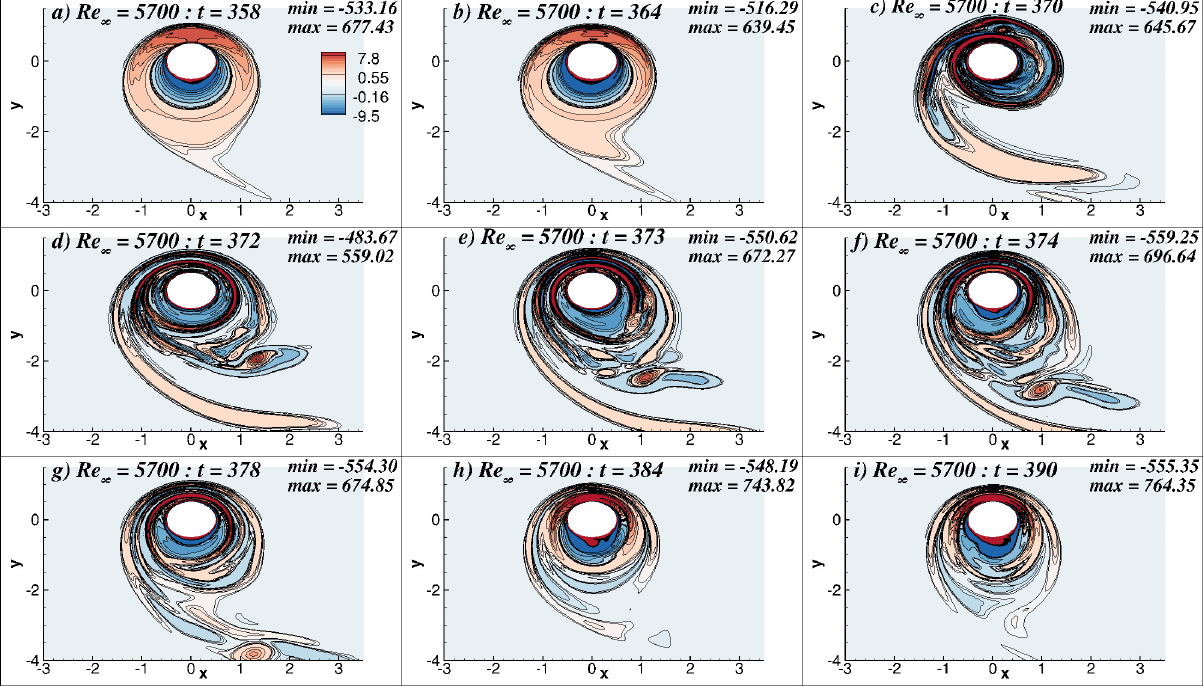}
%% Use \caption command for figure caption and label.
\caption{Spanwise vorticity contours for $U^*_s = 10U_{\infty}$ and super-critical $Re_{\infty} = 5700$, shown at indicated times starting from the onset of instability.}
\label{fig14}
\end{figure*}

The vorticity dynamics for the supercritical $Re_{\infty}$ of 5700 is shown in Fig. \ref{fig14}, starting from the onset time of the instability. Here, the vortical disturbances evolve much faster than that in the sub-critical and critical flow fields. The vortices in the wake of the cylinder are distorted, stretched, and irregular, but shed periodically. With further increase in $Re_{\infty}$, secondary instabilities are observed with a transition to quasi-periodic and chaotic wake. In the near-wake region, asymmetric, alternating vortex patterns are observed. Stronger vorticity roll-ups are observed such as in Fig. \ref{fig14}(c) at earlier time-instants compared to the critical $Re_{\infty}$ shown in Fig. \ref{fig13}. The vortical structures are more intense and highly localized structures with rotation-induced asymmetry. For higher $Re_{\infty}$ of 6000 plotted in Fig. \ref{fig5}, there is presence of multi-scale, irregular vortical structures. The coherent vortical structures are unsteady and evolving, indicating nonlinear interactions post-bifurcation eventually leading to chaos \citep{sipp2007global}.

\begin{figure*}%% placement specifier
\centering
\includegraphics[width=.9\textwidth]{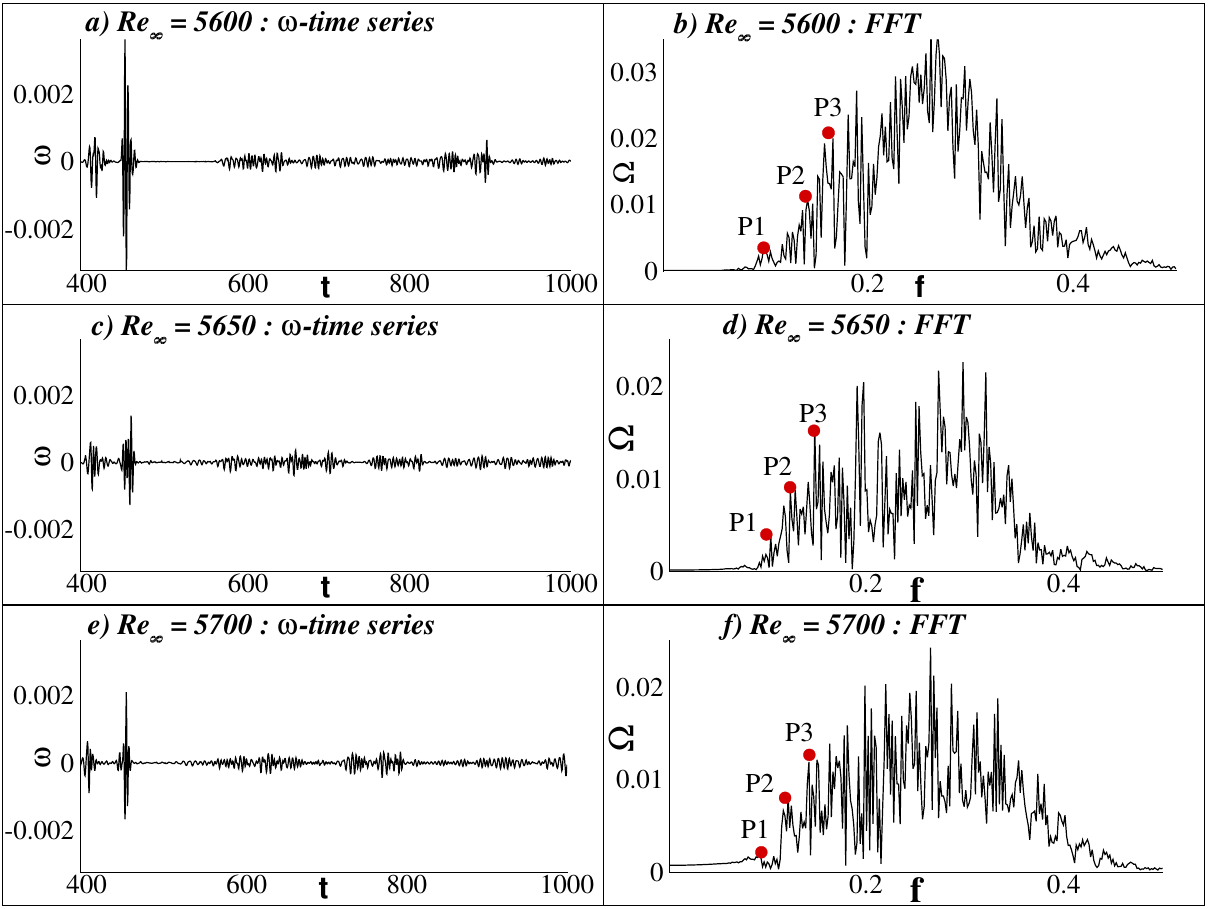}
%% Use \caption command for figure caption and label.
\caption{Time-series of spanwise vorticity probed at $x = 0, y = -1$ for (a) $Re_{\infty} = 5600$, (c) $Re_{\infty} = 5650$, and (e) $Re_{\infty} = 5700$, truncated to only show the growth of disturbance. The corresponding spectra are shown in frames (b), (d), and (f), respectively to explain the bifurcation point in Fig. \ref{fig11}.}
\label{fig15}
\end{figure*}

The time evolution of spanwise vorticity and corresponding FFT are shown in Fig. \ref{fig15} for sub-critical $Re_{\infty} = 5600$, critical $Re_{\infty} = 5650$, and super-critical $Re_{\infty} = 5700$, starting from the onset time of instability. For the sub-critical $Re_{\infty}$ shown in Fig. \ref{fig12}, flow is found to be weakly unsteady (near global stability limit). There is no vortex shedding, but quasi-periodic, low-energy transient fluctuations occur. The rotation suppresses shear-layer instabilities, but residual motion from long-time transients persist. Peaks are noted in the frequency plane in Fig. \ref{fig15}(b), marked as P1, P2 and P3, which are tabulated in Table \ref{tab3}. The first peak corresponds to the low-frequency global mode, while the second peak is due to a weak interaction mode between vorticity generated on the top and bottom surfaces of the cylinder. The third frequency arises from nonlinear interactions between the vortical structures generated in the wake of the cylinder. These peaks retain their identity in the vorticity field and are slow-decaying, so their spectral power is concentrated leading to higher amplitude, compared to the critical $Re_{\infty} = 5650$. The sub-critical flow supports slowly oscillating, quasi-steady vortical structures. The vorticity distribution for critical $Re_{\infty}$, shown in Fig. \ref{fig15}(c), undergoes a Hopf bifurcation and the flow becomes unstable to a time-periodic oscillation. This induces the Fourier amplitude to be distributed over the high frequency range as shown in Fig. \ref{fig15}(d), with the three identified peaks (P1-P3) having comparable amplitude to the subsequent ones. The newly emergent oscillatory mode is typically faster than the low-frequency transients of the sub-critical regime. This is the fundamental frequency of the bifurcated solution (i.e., first harmonic from linear instability), recorded in Table \ref{tab3}. At this stage, the system exhibits near-linear behavior, and nonlinear energy transfer is limited leading to lower Fourier amplitudes compared to sub-critical case. Flow is unsteady, with lower coherent vorticity strength. The post-bifurcation regime, shown in Fig. \ref{fig15}(e), has fully established quasi-periodic vortical structures in the wake of the cylinder. The wake becomes nonlinear, and multiple modes can interact and couple. Three peaks are noted in the frequency plane in Fig. \ref{fig15}(f) which correspond to the (i) fundamental frequency of the global mode, i.e. P1 in Table \ref{tab3} due to rotational asymmetry, (ii) a secondary mode due to nonlinear interaction between global instability mode and its superharmonic (i.e. $P2(f) = 1.25 \times P1(f)$). From Table \ref{tab3}, it is observed that the last peak in the frequency plane is such that $P3(f) = 1.45 \times P1(f)$. In the supercritical regime, the flow has stronger, coherent structures which generate consistent, high-energy periodic signals over a wide range of spatial scales, contributing to weaker Fourier amplitudes in the spectrum.

\begin{table}
\begin{center}
\def~{\hphantom{0}}
\caption{Dominant frequencies and their amplitudes, marked as P1, P2, P3 in Fig. \ref{fig15}}
\begin{tabular}{lccccccc}
 \hline
$Re_{\infty}$ & P1($f$) & P1($\Omega$) & P2($f$) & P2($\Omega$) & P3($f$) & P3($\Omega$)  \\ [3pt] 
 \hline 
 5600 & 0.1010 & 0.00273 & 0.14145 & 0.01072 & 0.166118 & 0.02001 \\ 
 \hline
 5650 & 0.1038 & 0.00389 & 0.12355 & 0.00872 & 0.14827 & 0.01572 \\
 \hline
 5700 & 0.0993 & 0.00167 & 0.12135 & 0.0080 & 0.14239 & 0.011897 \\
 \hline
 \end{tabular}
\label{tab3}
\end{center}
\end{table}

To visualize the frequency results of Table \ref{tab3}, we have added three supplementary videos for sub-critical $Re_{\infty} = 5600$, critical $Re_{\infty} = 5650$, and super-critical $Re_{\infty} = 5700$. In the top frame, the time-series of spanwise vorticity is depicted and, in the bottom frame, contour plots for spanwise vorticity are depicted. Each time-period corresponding to the frequency P1 of Table \ref{tab3} is marked in the video as $1TP, 2TP, ..., 10TP$. The onset of instability (OT) and various equilibrium states (marked in time-series as P and Q) are also shown in the video. The complementary flow field contour plot explains how shedding process changes for sub-critical to super-critical $Re_{\infty}$. The supplementary animations over ten time-periods of the first frequencies (P1) of Table \ref{tab3} are uploaded for $Re_{\infty}$ = 5600, 5650, and 5700, as mentioned in the data availability statement.

\subsection{Role of compressibility at a high rotation rate}

All viscous flows exhibit rotationality, quantified via the vorticity vector, $\vec{\omega}$. Enstrophy, defined as the dot product of the vorticity vector $\vec{\omega}$ with itself, $\Omega = \vec{\omega}\cdot \vec{\omega}$, measures the intensity of rotational motion in the flow, and is directly linked to turbulent dissipation and instability growth \cite{sengupta2018enstrophy}. The transport equation of compressible enstrophy \cite{suman2022novel} serves as a valuable tool for analyzing the generation, distribution, and evolution of enstrophy during transition to turbulence in various internal and external flows. The formulation of the compressible enstrophy transport equation (CETE) from the compressible NSE was introduced in an earlier work \cite{suman2022novel}. For finite free-stream $M_{\infty}$ and high rotation rates such as those reported here, the density variations, dilatation, and baroclinic torque terms become non-negligible. These effects directly modify vorticity generation, dissipation, and redistribution, which cannot be captured by the incompressible enstrophy equation. The CETE provides the only way to correctly quantify how vorticity is produced by both shear and density gradients (baroclinic effects), how it is amplified or damped by compressibility-related dilatation, and how the rotation-induced pressure and density fields couple to vorticity dynamics in the boundary layer and wake. Application of CETE to hydrodynamic instabilities \cite{joshi2025comparing, joshi2024highly, sengupta2023effects} has demonstrated a dominant role of viscous terms in enstrophy evolution for buoyancy-dominated flows. For advection-dominated flows, however, the vortex stretching term emerges as a significant contributor to enstrophy evolution alongside viscous effects. Enstrophy growth often traces back to baroclinic contributions, especially once quasi-periodic coherent vortical structures are observed in the flow field. For external aerodynamics \cite{sengupta2023compressibility, sengupta2025compressible}, on the other hand, the maximum contribution to enstrophy arises from the vortex stretching term and the viscous stress terms.

This section applies CETE to 2D simulations of the compressible flow past a rotating cylinder for $Re_{\infty}$ in the range of 1000 to 6000. The objective is to investigate the enstrophy dynamics governing the relative contributions of various physical mechanisms to the growth of instabilities governed by the Magnus-Robins effect, which is a first study of its kind. The constituent terms of the CETE are as follows \cite{suman2022novel}:

	\begin{equation}
		\begin{aligned}
			\frac{D{\Omega }}{Dt}
			= & \; 2\vec{\omega } \cdot \left[(\vec{\omega} \cdot {\nabla}) \vec{V}\right] - 2({\nabla} \cdot \vec{V}) \Omega \\
			& + \left(\frac{2}{\rho^{2}}\right) \vec{\omega } \cdot \left[\left({\nabla \rho} \times {{\nabla p}}\right)\right] -\left(\frac{2}{\rho^{2}}\right)\vec{\omega } \cdot \left[{\nabla \rho} \times {\nabla} \left(\lambda ({\nabla} \cdot \vec{V})\right)\right] \\
			& +\left(\frac{4}{\rho}\right)\vec{\omega } \cdot \left[{\nabla} \times \left[{\nabla} \cdot \left(\mu S \right)\right]\right] -\left(\frac{4}{\rho^2}\right)\vec{\omega } \cdot \left({\nabla \rho} \times ({\nabla} \cdot \left(\mu S\right)) \right)
		\end{aligned}
		\label{CETE}
	\end{equation}
	
The various terms of Eq. \eqref{CETE} are as follows:%\\[1.5ex]

\begin{itemize}
		
\item $2\vec{\omega } \cdot \left[(\vec{\omega} \cdot {\nabla}) \vec{V}\right]$ : Contribution to enstrophy due to vortex stretching (T1). %\\[1.25ex]
		
\item $({\nabla} \cdot \vec{V}) \Omega$: Enstrophy growth/decay due to compressibility (T2). %\\[1.25ex]
		
\item $\left(\frac{1}{\rho^{2}}\right) \vec{\omega } \cdot \left[\left({\nabla \rho} \times {{\nabla p}}\right)\right]$: Contribution from baroclinic term due to misalignment of gradients of pressure and density (T3). %\\[1.25ex]
		
\item $\left(\frac{1}{\rho^{2}}\right)\vec{\omega } \cdot \left[{\nabla \rho} \times {\nabla} \left(\lambda ({\nabla} \cdot \vec{V})\right)\right]$ : Contribution due to misalignment of vorticity and bulk viscosity gradients (T4).%\\[1.25ex]
		
\item $\left(\frac{1}{\rho}\right)\vec{\omega } \cdot \left[{\nabla} \times \left[{\nabla} \cdot \left(\mu S \right)\right]\right]$: Diffusion of enstrophy due to viscous action (T5).%\\[1.25ex]
		
\item $\left(\frac{1}{\rho^2}\right)\vec{\omega } \cdot \left({\nabla \rho} \times ({\nabla} \cdot \left(\mu S\right)) \right)$: Contribution due to misalignment of gradients of density and divergence of viscous stresses (T6).%\\[1.25ex]
\end{itemize}

\begin{figure*}%% placement specifier
\centering
\includegraphics[width=.9\textwidth]{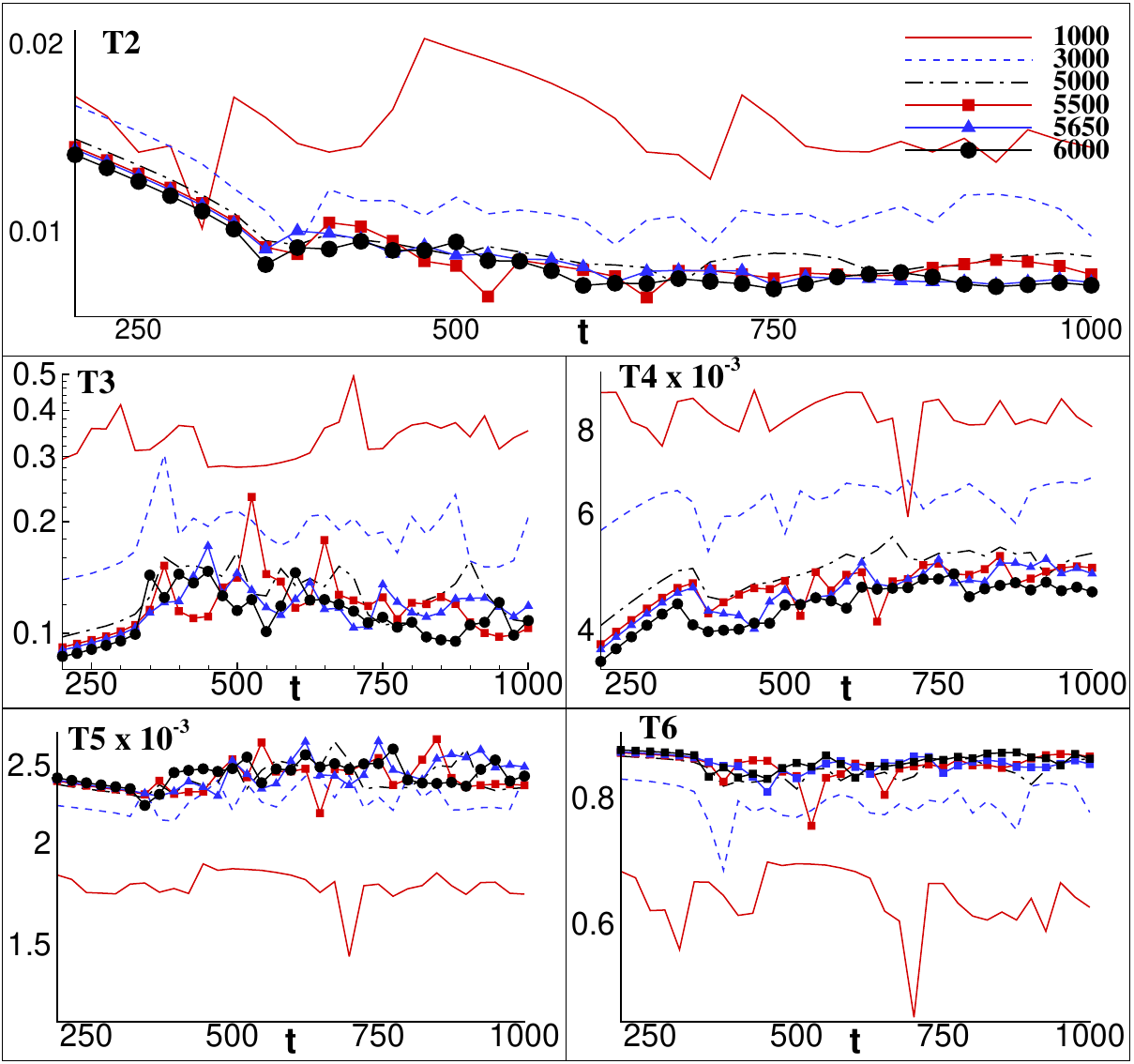}
%% Use \caption command for figure caption and label.
\caption{Evolution of normalized CETE budget terms for $Re_{\infty}$ = 1000, 3000, 5000, 5500, 5650, and 6000 to show relative contributions to enstrophy growth.}
\label{fig16}
\end{figure*}

Figure \ref{fig16} shows the evolution of the relative contributions of terms T2-T6 of Eq. \eqref{CETE} for the indicated $Re_{\infty}$. As the flow is 2D, the vortex stretching term, T1 is notably absent. Term T2, which attributes enstrophy growth due to compressibility/dilatation effects, shows a reduction with increase in $Re_{\infty}$ but beyond $Re_{\infty} > 5000$, T2 achieves a higher relative contribution with further increment in $Re_{\infty}$. At low to moderate $Re_{\infty}$, increasing $Re_{\infty}$ tends to make the flow more inertial and reduce large-scale divergence effects relative to shear-driven vorticity production, so the relative importance of the dilatation contribution (T2) falls. For $Re_{\infty} > 5000$, in the wake of the cylinder, local boundary-layer transition occurs which induces stronger small-scale gradients, and much larger local tangential speeds. This effect is more pronounced with rotation - which increases local Mach number, compression/expansion events and baroclinic coupling (noted by higher T3 values for $Re_{\infty} > 5000$). These cause $\nabla \cdot \vec{V}$ and its correlation with enstrophy to grow, so relative contribution of T2 rises again. Overall, the largest contribution is made by viscous stress term, T6 followed by the baroclinic term, T3. Compressibility effects quantified by T2 is the third largest contributor, and thus, the incorporation of a compressible formulation is necessary.

\subsection{Data-driven modelling using artificial neural network}

The onset of flow instability, lift and drag measurements are central to the design and control of bluff bodies across engineering applications (from wind turbines and bridges to micro-aero devices). Traditional high-fidelity numerical simulations and reduced-order models are computationally expensive to run repeatedly across multi-parameter design spaces. These may fail to capture strongly nonlinear transitions near bifurcation thresholds. A data-driven surrogate, specifically an artificial neural network (ANN) — offers a complementary capability: a compact, low-latency mapping from non-dimensional inputs like $Re_{\infty}$ to quantities of interest (onset of instability, $C_l$, $C_d$). This in turn enables fast parametric scans, uncertainty quantification, and real-time control and inverse design of systems. Additionally, ANNs can learn strongly nonlinear relationships, for instance, near Hopf bifurcations shown in Fig. \ref{fig11} or transitions caused by rotation that linear surrogates miss \cite{san2018machine}. Once trained, the ANN can interpolate between simulated points more robustly than simple regression. The framework can be extended to higher-dimensional parameter spaces (by inclusion of Mach number and rotation rate) using slight modifications in feature engineering and weight bias.

The ANN is a computational model consisting of interconnected artificial neurons organized into an input layer, one or more hidden layers, and an output layer. The hidden layers contain weights and biases that modify the strength of signal passing from one layer to the next, either amplifying or attenuating them. These are particularly useful for modelling complex multiphysical systems with multivariate properties, which is gaining significant importance in modelling complex fluid dynamics problem \citep{bengio2017deep}. In this study, we use the ANN to analyze 101 data points generated through simulations, to develop a model to predict different critical parameters of compressible flow past rotating cylinder. In this ANN model, we use a deep learning neural network with 12 hidden layers consisting of 256, 256, 128, 128, 64, 64, 32,32, 16, 16, 8, and 8 neurons, respectively. Preliminary tuning of the ANN involved tests with 4, 6, and 8 hidden layers but the accuracy achieved was not satisfactory. For 14 hidden layers also, the model accuracy decreased. The best accuracy was obtained with 12 layers with each layers having progressively decreased number of neurons in each layer, as explained earlier. This helps in balancing model capacity with regularization. For the ANN, we use exponential linear unit (ELU) as the activation function as the parameters studied (maximum $C_l$, maximum $C_d$, and onset time of instability) are highly nonlinear functions of the input parameter, $Re_{\infty}$. The ELU has the specialty of operating well for nonlinear functional relationships and also helps to mitigate the issue of vanishing gradients. For optimization, the ANN uses the Adam optimizer with default learning rate and Huber loss function was selected during training stage for its resilience to outliers. The high-fidelity simulation for each parameter requires 9901 core hours whereas, once trained, the ANN evaluates in 0.00006\% of the time, enabling dense parametric sweeps.

In the ANN model, key features are engineered from the functional relationships between $Re_{\infty}$ and the three output parameters viz. maximum $C_l$, maximum $C_d$, and onset time of instability, reported in Figs. \ref{fig10} and \ref{fig11}. Depending on the input $Re_{\infty}$, the functional relationship may follow a linear, quadratic, cubic, logarithmic or a reciprocal function. This choice is made on the basis of the flow physics inferred in Figs. \ref{fig10} and \ref{fig11}. For example, linear function is adopted when there is direct relationship between $Re_{\infty}$ and the output flow parameter. The quadratic term, on the other hand, is chosen when the secondary instabilities in the transitional flow dominate and often exhibit quadratic growth. Similarly, when compressibility effects are important in the flow, a cubic distribution is often adopted. The logarithmic term is used to satisfy log-law velocity profile in turbulent boundary layer and reciprocal function is used to provide viscous correction. This enhances the input space and capture complex nonlinear relationships between $Re_{\infty}$ and critical flow parameters such as onset time of instability, maximum $C_l$ and maximum $C_d$. To design the ANN models, linear, quadratic and cubic distributions are considered for all three output parameters, while for maximum $C_l$ versus $Re_{\infty}$, we use an additional logarithmic function. Similarly, for maximum $C_d$ versus $Re_{\infty}$, an additional reciprocal function is used. To emphasize the model’s predictive capability in high $Re_{\infty}$-regimes, custom sample weights are assigned, enhancing the contribution of data samples in the critical parts of the distributions where fluctuations are significant.

The ANN is trained using randomly selected data sets from the 101 two-dimensional simulations to prevent overfitting. Three training scenarios were tested using 80\%, 85\% and 90\% of the data. Data sets were split into three subsets: training set, testing set and validation set respectively to mitigate the risk of overfitting of the model. The dataset is split into training sets (80\%, 85\%, or 90\%), and out of the remaining data (20\%, 15\%, and 10\%), we further split into 10\% for validation and 90\% for testing. To quantify the performance of the ANN model, we recorded the mean absolute error (MAE), mean square error (MSE), and the accuracy in Tables \ref{tab4} to \ref{tab6}, for maximum $C_d$, maximum $C_l$, and onset time, respectively. For all the test runs, the number of epochs is varied in intervals of 100. The python script used for the ANN is provided as supplementary data.

For the maximum $C_d$ versus $Re_{\infty}$ distribution shown in Fig. \ref{fig10}(b), the batch size is chosen as 20. The maximum accuracy for 80\%, 85\%, and 90\% training data is obtained as 78.8\%, 81.5\%, and 90.03\%, respectively in Table \ref{tab4}. For the maximum $C_l$ versus $Re_{\infty}$ distribution shown in Fig. \ref{fig10}(a), the batch size is chosen as 15. The maximum accuracy for 80\%, 85\%, and 90\% training data is obtained as 99.3\%, 99.83\%, and 99.8\%, respectively in Table \ref{tab5}. For $C_l$, the results do not vary much between the various training datasets used, which indicates that its variation with $Re_{\infty}$ is relatively straightforward to model compared to $C_d$. This is affirmed by the gradual dip in the $C_l$ distribution in Fig. \ref{fig10}(a), wherein small-amplitude fluctuations appear for $Re_{\infty} > 5650$. In contrast, the $C_d$ distribution in Fig. \ref{fig10}(b), shows a different relation with $Re_{\infty}$ depending on the range. The fluctuations for $Re_{\infty} > 5650$ in the $C_d$ distribution are much steeper than that for $C_l$, which makes it harder to model using the ANN. For the onset time of instability versus $Re_{\infty}$ distribution shown in Fig. \ref{fig11}, the ANN model uses a batch size of 10. The maximum accuracy for 80\%, 85\%, and 90\% training data is obtained as 99.8\%, 99.94\%, and 99.97\%, respectively in Table \ref{tab6}. The variation of onset time with $Re_{\infty}$ also follows a simple functional relationship, for which the ANN does not vary much between the three training data sets. This suggests that the level of refinement offered by the 101 data sets is not necessary for $C_l$ and onset time of instability, but for $C_d$ where there are significant variations with $Re_{\infty}$, a dense data set is paramount to the success of the ANN model.

\begin{table}
\begin{center}
\def~{\hphantom{0}}
\caption{Parameter used in ANN, MAE, MSE and Accuracy in case of $Re_{\infty}$ vs Max $C_d$}
\begin{tabular}{lccccc}
 \hline
 Training Data & Epochs & MAE & MSE & Accuracy(\%) \\ [3pt]
 \hline
 
 80\% & 100 & 0.4658 & 1.3518 & 74.27 \\
 \hline
 80\% & 200 & 0.5021 & 1.1920 & 77.31 \\
 \hline
 80\% & 300 & 0.5115 & 1.4562 & 72.28 \\
 \hline
 80\% & 400 & 0.5060 & 1.3549 & 74.21 \\
 \hline
 {\bf 80\%} & {\bf 500} & {\bf 0.4473} & {\bf 1.1138} & {\bf 78.80} \\
 \hline
 80\% & 600 & 0.5371 & 1.5454 & 70.58 \\
 \hline
 85\% & 100 & 0.4616 & 1.5711 & 76.37 \\
 \hline
 85\% & 200 & 0.4188 & 1.2914 & 80.58 \\
 \hline
 85\% & 300 & 0.4924 & 1.4693 & 77.90 \\
 \hline
 {\bf 85\%} & {\bf 400} & {\bf 0.4919} & {\bf 1.2286} & {\bf 81.52} \\
 \hline
 85\% & 500 & 0.6252 & 2.0898 & 68.57 \\
 \hline
 85\% & 600 & 0.4963 & 1.9580 & 70.55 \\
 \hline
 90\% & 100 & 0.4692 & 1.6481 & 79.29 \\
 \hline
 90\% & 200 & 0.4421 & 1.6197 & 79.65 \\
 \hline
 90\% & 300 & 0.4959 & 1.9651 & 75.31 \\
 \hline
 90\% & 400 & 0.4145 & 1.4783 & 81.42 \\
 \hline
 90\% & 500 & 0.5187 & 1.2285 & 84.56 \\
 \hline
  \textbf{90\%} & \textbf{600} & \textbf{0.2892} & \textbf{0.7937} & \textbf{90.03} \\
 \end{tabular}
\label{tab4}
\end{center}
\end{table}

\begin{table}
\begin{center}
\def~{\hphantom{0}}
\caption{Parameter used in ANN, MAE, MSE and Accuracy in case of $Re_{\infty}$ vs Max $C_l$}
\begin{tabular}{lccccc}
 \hline
 Training Data & Epochs & MAE & MSE & Accuracy(\%) \\ [3pt]
 \hline
 
 80\% & 100 & 0.1752 & 0.0752 & 99.06 \\
 \hline
 {\bf 80\%} & {\bf 200} & {\bf 0.1293} & {\bf 0.0579} & {\bf 99.28} \\
 \hline
 80\% & 300 & 0.1373 & 0.0709 & 99.11 \\
 \hline
 80\% & 400 & 0.1505 & 0.0628 & 99.22 \\
 \hline
 80\% & 500 & 0.1222 & 0.0676 & 99.16 \\
 \hline
 80\% & 600 & 0.1649 & 0.0985 & 98.77 \\
 \hline
 85\% & 100 & 0.1131 & 0.0257 & 99.69 \\
 \hline
  \textbf{85\%} & \textbf{200} & \textbf{0.0835} & \textbf{0.0139} & \textbf{99.83} \\
 \hline
 85\% & 300 & 0.0780 & 0.0182 & 99.78 \\
 \hline
 85\% & 400 & 0.0969 & 0.0225 & 99.73 \\
 \hline
 85\% & 500 & 0.0856 & 0.0156 & 99.81 \\
 \hline
 85\% & 600 & 0.1058 & 0.0256 & 99.69 \\
 \hline
 90\% & 100 & 0.0977 & 0.0316 & 99.57 \\
 \hline
 90\% & 200 & 0.1368 & 0.0253 & 99.65 \\
 \hline
 90\% & 300 & 0.1163 & 0.0337 & 99.54 \\
 \hline
 {\bf 90\%} & {\bf 400} & {\bf 0.0838} & {\bf 0.0146} & {\bf 99.80} \\
 \hline
 90\% & 500 & 0.0916 & 0.0154 & 99.79 \\
 \hline
 90\% & 600 & 0.0734 & 0.0156 & 99.79 \\
 \end{tabular}
\label{tab5}
\end{center}
\end{table}

\begin{figure*}%% placement specifier
\centering
\includegraphics[width=.95\textwidth]{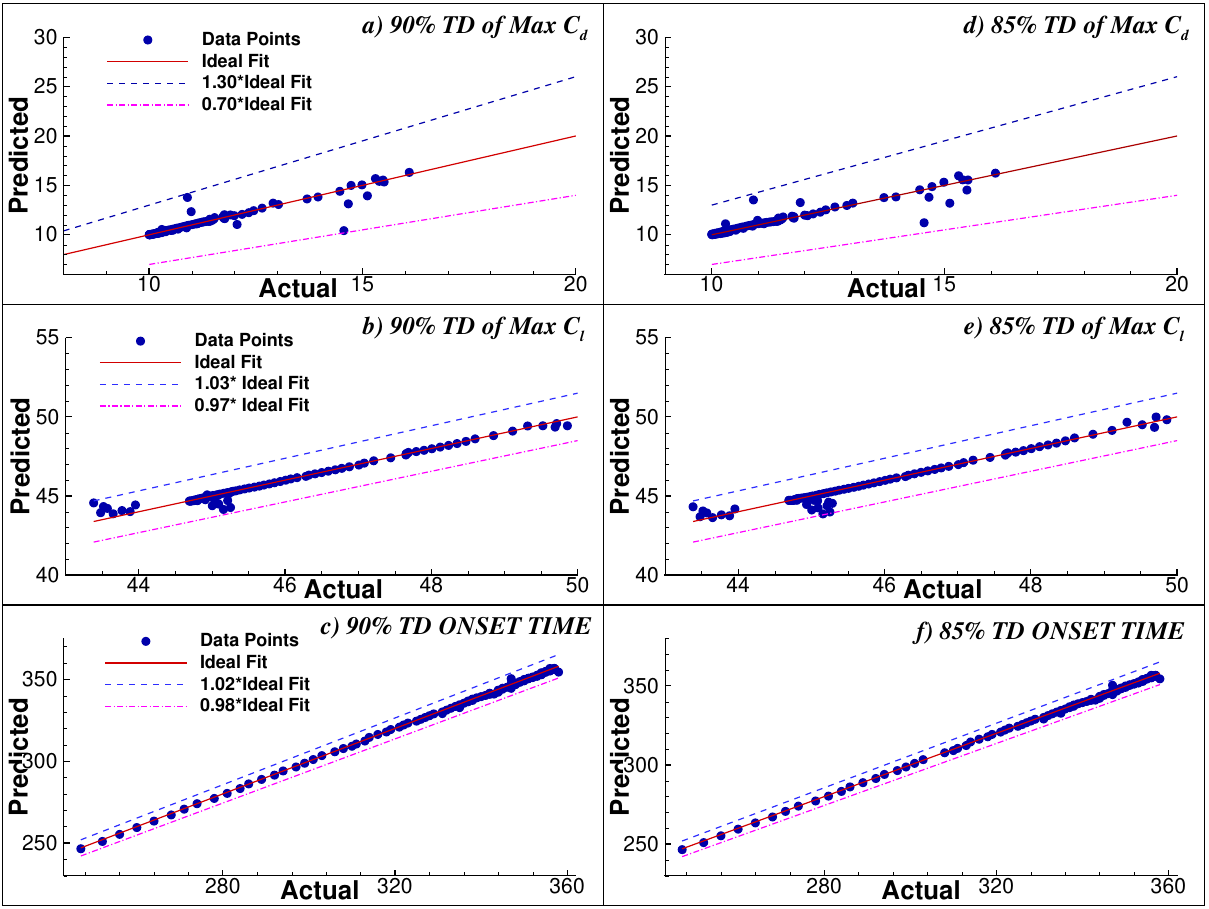}
%% Use \caption command for figure caption and label.
\caption{Parity plot of predicted value from ANN and actual value from simulation for (a)-(c) 90\% training data, and (d)-(f) 85\% training data. The maximum $C_d$, maximum $C_l$, and onset time are shown in frames (a), (d); (b), (e); and (c),(f), respectively.}
\label{fig17}
\end{figure*}

The parity plots for the variation of maximum $C_d$, maximum $C_l$, and onset time of instability with $Re_{\infty}$ are shown in Fig. \ref{fig17} for 85\% and 90\% training data in frames (a-c) and (d-f), respectively. This graph reveals that the predictions from ANN fall within 30\% of the actual values for maximum $C_d$, which has the most complex functional relationship with $Re_{\infty}$. For prediction of $C_d$ with 80\% and 85\% training data the agreement is poor beyond the point of bifurcation, which suggests that more refined simulation data is to be augmented near the bifurcation boundaries. Further, the result from the ANN could be improved by including a physics-informed residual term in the neural network. On the other hand, for maximum $C_l$ and onset time of instability, the ANN predictions are within 3\% and 2\% of the actual data points, respectively. There is not much difference between the two sets of results with 85\% and 90\% training data. 

%%%%%%%%%%%%Limitations %%%%%%%%%%%%%%%%%%%%

The ANN-based approach effectively captures the highly nonlinear dynamics inherent in the compressible flow past rotating cylinders, providing  accurate and generalizable predictions for three critical flow parameters. The methodological framework which encompasses feature engineering, data scaling, weighted training and careful model design, underscores the robustness and suitability of deep learning for modelling complex fluid dynamics. To be useful beyond interpolation, the ANN must be trained and validated with physics-guided constraints that promote robust generalization, such as wide parameter sampling, physics-informed loss terms, and explicit uncertainty estimates. ANN has limited extrapolation reliability, a known constraint of data-driven models. To address this, data sets must be kept aside for extrapolation test sets where $Re_{\infty}$ is outside the training range but rotation or geometry parameters lie within. The robustness of the ANN can also be tested for changes in $Re_{\infty}$ is altered by a different mechanism than in training (e.g., training with velocity-driven change, testing with viscosity-driven or diameter-driven change). When constructed and validated in this way, an ANN serves as a practical surrogate that accelerates design cycles, augments high-fidelity solvers, and guides experiments — especially in regimes where computing full simulations for every parameter combination is infeasible. The main practical value of the presented ANN lies in fast interpolation within the validated regime, enabling efficient flow-field characterization and parametric optimization, while the compressible flow solver remains available for reference in unexplored regimes.

\begin{table}
\begin{center}
\def~{\hphantom{0}}
\caption{Parameter used in ANN, MAE, MSE and Accuracy in case of $Re_{\infty}$ vs onset time of instability}
\begin{tabular}{lccccc}
 \hline
 Training Data & Epochs & MAE & MSE & Accuracy(\%) \\ [3pt]
 \hline
 
 80\% & 100 & 0.6745 & 1.8864 & 99.80 \\
 \hline
 80\% & 200 & 1.3421 & 5.9754 & 99.39 \\
 \hline
 80\% & 300 & 1.0469 & 3.8067 & 99.61 \\
 \hline
 80\% & 400 & 1.9391 & 10.9339 & 98.88 \\
 \hline
 80\% & 500 & 0.8000 & 3.9249 & 99.60 \\
 \hline
 80\% & 600 & 0.7017 & 2.1869 & 99.77 \\
 \hline
 85\% & 100 & 1.1892 & 1.7795 & 99.82 \\
 \hline
 85\% & 200 & 1.0162 & 1.5759 & 99.84 \\
 \hline
 85\% & 300 & 1.4377 & 2.5794 & 99.74 \\
 \hline
 85\% & 400 & 0.5879 & 0.6527 & 99.93 \\
 \hline
 85\% & 500 & 0.6053 & 0.5754 & 99.94 \\
 \hline
 85\% & 600 & 0.6216 & 0.5596 & 99.94 \\
 \hline
 90\% & 100 & 0.5451 & 0.4412 & 99.95 \\
 \hline
 90\% & 200 & 0.4723 & 0.3443 & 99.96 \\
 \hline
 90\% & 300 & 0.5218 & 0.4133 & 99.96 \\
 \hline
 90\% & 400 & 0.6363 & 0.6311 & 99.94 \\
 \hline
 90\% & 500 & 0.6213 & 0.5543 & 99.94 \\
 \hline
  \textbf{90\%} & \textbf{600} & \textbf{0.4551} & \textbf{0.2574} & \textbf{99.97} \\
  
 \end{tabular}
\label{tab6}
\end{center}
\end{table}

\section{Summary and conclusions}

The vorticity dynamics for a range of $Re_{\infty}$ are examined in Figs. \ref{fig3} to \ref{fig5}. The coherent vortical structures in the wake of the spinning cylinder are found to be a strong function of $Re_{\infty}$. For lower $Re_{\infty}$, these opposite signed vorticity appeared as \lq blobs' loosely surrounding the cylinder. As the $Re_{\infty}$ is increased progressively, negative and positive vorticity patches are wrapped around the cylinder as tightly wound spirals.  The temporal scale evolution shown in Figs. \ref{fig6} and \ref{fig7} and the recorded peaks in the frequency plane in Table \ref{tab1}, reveal that as $Re_{\infty}$ is increased, vorticity is redistributed along the bottom half of the rotating cylinder, and the time period of oscillations decreases as the Magnus-Robins instability appears.  For flow with $Re_{\infty}$ in the range of 5000-
5500, the compressibility effect dominates and inhibits mixing between shear layers, which leads to delayed vortex shedding.

The unsteady force distribution in Figs. \ref{fig8} to \ref{fig10} as a function of $Re_{\infty}$ shows the presence of certain critical $Re_{\infty}$ beyond which the dynamical system behavior bifurcates. Beyond a critical $Re_{\infty}$ of 5650, flow enters a fully unsteady transitional regime. It exhibits bifurcation and chaotic wake states, highly sensitive to small changes in $Re_{\infty}$. Drag and lift fluctuates due to changes in instantaneous pressure distribution, wake turbulence and bifurcation. The onset of the instability is tracked for various $Re_{\infty}$ in Fig. \ref{fig11}, with the initial trend showing an increase in onset time with $Re_{\infty}$ till $Re_{\infty} = 5650$. For high rotation rates, a strong circulation is induced, leading to
asymmetric pressure fields, suppressed separation and a stabilized wake near the bottom of the cylinder which delay the growth of disturbances in the shear layer, particularly at moderate $Re$. When $Re_{\infty} = 5650$ is attained, flow structures in the wake of the cylinder and in the shear layer are receptive and small disturbances grow spatio-temporally till the system becomes globally unstable.

The point of bifurcation, i.e. $Re_{\infty} = 5650$ is examined in detail in Figs. \ref{fig12} to \ref{fig14} by showing the vorticity dynamics for a sub-critical, critical and super-critical $Re_{\infty}$. The corresponding time-series and spectra are compared in Fig. \ref{fig15}. For the sub-critical case, flow is weakly unsteady (near global stability limit) with no vortex shedding. However, quasi-periodic, low-energy transient fluctuations occur as rotation suppresses shear-layer instabilities, but residual motion from long-time transients persist. At the critical $Re_{\infty}$ system undergoes a Hopf bifurcation and the flow becomes unstable to a time-periodic oscillation. The emergent oscillatory mode is faster than the low-frequency transients of
the sub-critical regime. The post-bifurcation regime, on the other hand, shows that the wake becomes nonlinear, and multiple modes can interact and couple. Three peaks are noted in the frequency plane (shown in Table \ref{tab3}) which correspond to fundamental frequency of the global mode, its superharmonic due to secondary instabilities in the shear layer, and secondary mode due to nonlinear interactions.

The application of an ANN to the plots of maximum $C_l$, $C_d$, and onset time with varying $Re_{\infty}$ in Figs. \ref{fig10} and \ref{fig11}, is shown in the parity plot between predicted ANN values and actual simulation values in Fig. \ref{fig17}. The associated error metrics are recorded in Tables \ref{tab4} to \ref{tab6}. The ANN provides a suitable alternative to the high-fidelity simulations reported here with accuracies in the range of 90\%-98\% for the considered parameters. The bifurcation analysis reported here serves as a valuable benchmarking database for compressible flow solvers and application of supervised machine learning methods. In the future, we intend to extend our analysis to report the role of rotation rates and Mach numbers for a fixed $Re_{\infty}$ in the compressible flow past a rotating cylinder.

\begin{acknowledgments}
The authors would like to acknowledge the use of high-performance computing facility CHANDRASEKHAR at Indian Institute of Technology Dhanbad for all the computations reported here. 
\end{acknowledgments}

\section*{Declaration of Interests}
The authors report no conflict of interest.

\section*{Data Availability}
The datasets produced from the high-fidelity simulations are available from the corresponding author upon reasonable request. A portion of the data for $Re$ = 1000, 2000, 3000, 4000, 5000, and 6000 is available at the following link: https://doi.org/10.5281/zenodo.17431094. Supplementary animations over ten time-periods of the significant flow events are uploaded for $Re_{\infty}$ = 5600, 5650, and 5700 on the following link: https://doi.org/10.5281/zenodo.17482560. Python codes for ANN are available at: https://doi.org/10.5281/zenodo.17482864.

%\appendix

%\section*{References}

\bibliography{rot_cyl}% Produces the bibliography via BibTeX.

\begin{thebibliography}{48}%
\makeatletter
\providecommand \@ifxundefined [1]{%
 \@ifx{#1\undefined}
}%
\providecommand \@ifnum [1]{%
 \ifnum #1\expandafter \@firstoftwo
 \else \expandafter \@secondoftwo
 \fi
}%
\providecommand \@ifx [1]{%
 \ifx #1\expandafter \@firstoftwo
 \else \expandafter \@secondoftwo
 \fi
}%
\providecommand \natexlab [1]{#1}%
\providecommand \enquote  [1]{``#1''}%
\providecommand \bibnamefont  [1]{#1}%
\providecommand \bibfnamefont [1]{#1}%
\providecommand \citenamefont [1]{#1}%
\providecommand \href@noop [0]{\@secondoftwo}%
\providecommand \href [0]{\begingroup \@sanitize@url \@href}%
\providecommand \@href[1]{\@@startlink{#1}\@@href}%
\providecommand \@@href[1]{\endgroup#1\@@endlink}%
\providecommand \@sanitize@url [0]{\catcode `\\12\catcode `\$12\catcode `\&12\catcode `\#12\catcode `\^12\catcode `\_12\catcode `\%12\relax}%
\providecommand \@@startlink[1]{}%
\providecommand \@@endlink[0]{}%
\providecommand \url  [0]{\begingroup\@sanitize@url \@url }%
\providecommand \@url [1]{\endgroup\@href {#1}{\urlprefix }}%
\providecommand \urlprefix  [0]{URL }%
\providecommand \Eprint [0]{\href }%
\providecommand \doibase [0]{http://dx.doi.org/}%
\providecommand \selectlanguage [0]{\@gobble}%
\providecommand \bibinfo  [0]{\@secondoftwo}%
\providecommand \bibfield  [0]{\@secondoftwo}%
\providecommand \translation [1]{[#1]}%
\providecommand \BibitemOpen [0]{}%
\providecommand \bibitemStop [0]{}%
\providecommand \bibitemNoStop [0]{.\EOS\space}%
\providecommand \EOS [0]{\spacefactor3000\relax}%
\providecommand \BibitemShut  [1]{\csname bibitem#1\endcsname}%
\let\auto@bib@innerbib\@empty
%</preamble>
\bibitem [{\citenamefont {White}(1994)}]{white1994fluid}%
  \BibitemOpen
  \bibfield  {author} {\bibinfo {author} {\bibfnamefont {F.~M.}\ \bibnamefont {White}},\ }\bibfield  {title} {\enquote {\bibinfo {title} {Fluid mechanics},}\ }\href@noop {} {\bibfield  {journal} {\bibinfo  {journal} {McGraw-Hill, New York}\ } (\bibinfo {year} {1994})}\BibitemShut {NoStop}%
\bibitem [{\citenamefont {Mittal}\ and\ \citenamefont {Kumar}(2003)}]{mittal2003flow}%
  \BibitemOpen
  \bibfield  {author} {\bibinfo {author} {\bibfnamefont {S.}~\bibnamefont {Mittal}}\ and\ \bibinfo {author} {\bibfnamefont {B.}~\bibnamefont {Kumar}},\ }\bibfield  {title} {\enquote {\bibinfo {title} {Flow past a rotating cylinder},}\ }\href@noop {} {\bibfield  {journal} {\bibinfo  {journal} {J. Fluid Mech.}\ }\textbf {\bibinfo {volume} {476}},\ \bibinfo {pages} {303--334} (\bibinfo {year} {2003})}\BibitemShut {NoStop}%
\bibitem [{\citenamefont {Sengupta}\ \emph {et~al.}(2003)\citenamefont {Sengupta}, \citenamefont {Kasliwal}, \citenamefont {De},\ and\ \citenamefont {Nair}}]{sengupta2003temporal}%
  \BibitemOpen
  \bibfield  {author} {\bibinfo {author} {\bibfnamefont {T.~K.}\ \bibnamefont {Sengupta}}, \bibinfo {author} {\bibfnamefont {A.}~\bibnamefont {Kasliwal}}, \bibinfo {author} {\bibfnamefont {S.}~\bibnamefont {De}}, \ and\ \bibinfo {author} {\bibfnamefont {M.}~\bibnamefont {Nair}},\ }\bibfield  {title} {\enquote {\bibinfo {title} {Temporal flow instability for magnus--robins effect at high rotation rates},}\ }\href@noop {} {\bibfield  {journal} {\bibinfo  {journal} {J. Fluids Struct.}\ }\textbf {\bibinfo {volume} {17}},\ \bibinfo {pages} {941--953} (\bibinfo {year} {2003})}\BibitemShut {NoStop}%
\bibitem [{\citenamefont {Stojković}, \citenamefont {Breuer},\ and\ \citenamefont {Durst}(2002)}]{stojkovic2002effect}%
  \BibitemOpen
  \bibfield  {author} {\bibinfo {author} {\bibfnamefont {D.}~\bibnamefont {Stojković}}, \bibinfo {author} {\bibfnamefont {M.}~\bibnamefont {Breuer}}, \ and\ \bibinfo {author} {\bibfnamefont {F.}~\bibnamefont {Durst}},\ }\bibfield  {title} {\enquote {\bibinfo {title} {Effect of high rotation rates on the laminar flow around a circular cylinder},}\ }\href@noop {} {\bibfield  {journal} {\bibinfo  {journal} {Phys. Fluids}\ }\textbf {\bibinfo {volume} {14}},\ \bibinfo {pages} {3160--3178} (\bibinfo {year} {2002})}\BibitemShut {NoStop}%
\bibitem [{\citenamefont {Glauert}(1957)}]{glauert1957flow}%
  \BibitemOpen
  \bibfield  {author} {\bibinfo {author} {\bibfnamefont {M.~B.}\ \bibnamefont {Glauert}},\ }\bibfield  {title} {\enquote {\bibinfo {title} {The flow past a rapidly rotating circular cylinder},}\ }\href@noop {} {\bibfield  {journal} {\bibinfo  {journal} {Proc. R. Soc. Lond. A Math. Phys. Sci.}\ }\textbf {\bibinfo {volume} {242}},\ \bibinfo {pages} {108--115} (\bibinfo {year} {1957})}\BibitemShut {NoStop}%
\bibitem [{\citenamefont {Teymourtash}\ and\ \citenamefont {Salimipour}(2017)}]{teymourtash2017compressibility}%
  \BibitemOpen
  \bibfield  {author} {\bibinfo {author} {\bibfnamefont {A.~R.}\ \bibnamefont {Teymourtash}}\ and\ \bibinfo {author} {\bibfnamefont {S.~E.}\ \bibnamefont {Salimipour}},\ }\bibfield  {title} {\enquote {\bibinfo {title} {Compressibility effects on the flow past a rotating cylinder},}\ }\href@noop {} {\bibfield  {journal} {\bibinfo  {journal} {Phys. Fluids}\ }\textbf {\bibinfo {volume} {29}} (\bibinfo {year} {2017})}\BibitemShut {NoStop}%
\bibitem [{\citenamefont {Sundaram}\ \emph {et~al.}(2021)\citenamefont {Sundaram}, \citenamefont {Sengupta}, \citenamefont {Sengupta},\ and\ \citenamefont {Suman}}]{sundaram2021multiscale}%
  \BibitemOpen
  \bibfield  {author} {\bibinfo {author} {\bibfnamefont {P.}~\bibnamefont {Sundaram}}, \bibinfo {author} {\bibfnamefont {T.~K.}\ \bibnamefont {Sengupta}}, \bibinfo {author} {\bibfnamefont {A.}~\bibnamefont {Sengupta}}, \ and\ \bibinfo {author} {\bibfnamefont {V.~K.}\ \bibnamefont {Suman}},\ }\bibfield  {title} {\enquote {\bibinfo {title} {Multiscale instabilities of magnus--robins effect for compressible flow past rotating cylinder},}\ }\href@noop {} {\bibfield  {journal} {\bibinfo  {journal} {Phys. Fluids}\ }\textbf {\bibinfo {volume} {33}} (\bibinfo {year} {2021})}\BibitemShut {NoStop}%
\bibitem [{\citenamefont {Suman}\ \emph {et~al.}(2022)\citenamefont {Suman}, \citenamefont {Sundaram}, \citenamefont {Puttam}, \citenamefont {Sengupta},\ and\ \citenamefont {Sengupta}}]{suman2022novel}%
  \BibitemOpen
  \bibfield  {author} {\bibinfo {author} {\bibfnamefont {V.~K.}\ \bibnamefont {Suman}}, \bibinfo {author} {\bibfnamefont {P.}~\bibnamefont {Sundaram}}, \bibinfo {author} {\bibfnamefont {J.~K.}\ \bibnamefont {Puttam}}, \bibinfo {author} {\bibfnamefont {A.}~\bibnamefont {Sengupta}}, \ and\ \bibinfo {author} {\bibfnamefont {T.~K.}\ \bibnamefont {Sengupta}},\ }\bibfield  {title} {\enquote {\bibinfo {title} {A novel compressible enstrophy transport equation-based analysis of instability during magnus--robins effects for high rotation rates},}\ }\href@noop {} {\bibfield  {journal} {\bibinfo  {journal} {Phys. Fluids}\ }\textbf {\bibinfo {volume} {34}} (\bibinfo {year} {2022})}\BibitemShut {NoStop}%
\bibitem [{\citenamefont {Prandtl}(1926)}]{prandtl1926application}%
  \BibitemOpen
  \bibfield  {author} {\bibinfo {author} {\bibfnamefont {L.}~\bibnamefont {Prandtl}},\ }\href@noop {} {\enquote {\bibinfo {title} {Application of the ``magnus effect'' to the wind propulsion of ships},}\ }\bibinfo {type} {Tech. Rep.}\ (\bibinfo {year} {1926})\BibitemShut {NoStop}%
\bibitem [{\citenamefont {Tokumaru}\ and\ \citenamefont {Dimotakis}(1993)}]{tokumaru1993lift}%
  \BibitemOpen
  \bibfield  {author} {\bibinfo {author} {\bibfnamefont {P.~T.}\ \bibnamefont {Tokumaru}}\ and\ \bibinfo {author} {\bibfnamefont {P.~E.}\ \bibnamefont {Dimotakis}},\ }\bibfield  {title} {\enquote {\bibinfo {title} {The lift of a cylinder executing rotary motions in a uniform flow},}\ }\href@noop {} {\bibfield  {journal} {\bibinfo  {journal} {J. Fluid Mech.}\ }\textbf {\bibinfo {volume} {255}},\ \bibinfo {pages} {1--10} (\bibinfo {year} {1993})}\BibitemShut {NoStop}%
\bibitem [{\citenamefont {Pralits}, \citenamefont {Brandt},\ and\ \citenamefont {Giannetti}(2010)}]{pralits2010instability}%
  \BibitemOpen
  \bibfield  {author} {\bibinfo {author} {\bibfnamefont {J.~O.}\ \bibnamefont {Pralits}}, \bibinfo {author} {\bibfnamefont {L.}~\bibnamefont {Brandt}}, \ and\ \bibinfo {author} {\bibfnamefont {F.}~\bibnamefont {Giannetti}},\ }\bibfield  {title} {\enquote {\bibinfo {title} {Instability and sensitivity of the flow around a rotating circular cylinder},}\ }\href@noop {} {\bibfield  {journal} {\bibinfo  {journal} {J. Fluid Mech.}\ }\textbf {\bibinfo {volume} {650}},\ \bibinfo {pages} {513--536} (\bibinfo {year} {2010})}\BibitemShut {NoStop}%
\bibitem [{\citenamefont {Salimipour}\ and\ \citenamefont {Anbarsooz}(2019)}]{salimipour2019surface}%
  \BibitemOpen
  \bibfield  {author} {\bibinfo {author} {\bibfnamefont {E.}~\bibnamefont {Salimipour}}\ and\ \bibinfo {author} {\bibfnamefont {M.}~\bibnamefont {Anbarsooz}},\ }\bibfield  {title} {\enquote {\bibinfo {title} {Surface temperature effects on the compressible flow past a rotating cylinder},}\ }\href@noop {} {\bibfield  {journal} {\bibinfo  {journal} {Phys. Fluids}\ }\textbf {\bibinfo {volume} {31}} (\bibinfo {year} {2019})}\BibitemShut {NoStop}%
\bibitem [{\citenamefont {Nagata}\ \emph {et~al.}(2018)\citenamefont {Nagata}, \citenamefont {Nonomura}, \citenamefont {Takahashi}, \citenamefont {Mizuno},\ and\ \citenamefont {Fukuda}}]{nagata2018direct}%
  \BibitemOpen
  \bibfield  {author} {\bibinfo {author} {\bibfnamefont {T.}~\bibnamefont {Nagata}}, \bibinfo {author} {\bibfnamefont {T.}~\bibnamefont {Nonomura}}, \bibinfo {author} {\bibfnamefont {S.}~\bibnamefont {Takahashi}}, \bibinfo {author} {\bibfnamefont {Y.}~\bibnamefont {Mizuno}}, \ and\ \bibinfo {author} {\bibfnamefont {K.}~\bibnamefont {Fukuda}},\ }\bibfield  {title} {\enquote {\bibinfo {title} {Direct numerical simulation of flow past a transversely rotating sphere up to a reynolds number of 300 in compressible flow},}\ }\href@noop {} {\bibfield  {journal} {\bibinfo  {journal} {J. Fluid Mech.}\ }\textbf {\bibinfo {volume} {857}},\ \bibinfo {pages} {878--906} (\bibinfo {year} {2018})}\BibitemShut {NoStop}%
\bibitem [{\citenamefont {Kundu}, \citenamefont {Cohen},\ and\ \citenamefont {Dowling}(2012)}]{kundu2012fluid}%
  \BibitemOpen
  \bibfield  {author} {\bibinfo {author} {\bibfnamefont {P.~K.}\ \bibnamefont {Kundu}}, \bibinfo {author} {\bibfnamefont {I.~M.}\ \bibnamefont {Cohen}}, \ and\ \bibinfo {author} {\bibfnamefont {D.~R.}\ \bibnamefont {Dowling}},\ }\href@noop {} {\emph {\bibinfo {title} {Fluid mechanics, 5th edition}}}\ (\bibinfo  {publisher} {Academic Press, Berlin},\ \bibinfo {year} {2012})\BibitemShut {NoStop}%
\bibitem [{\citenamefont {Nishioka}\ and\ \citenamefont {Sato}(1978)}]{nishioka1978mechanism}%
  \BibitemOpen
  \bibfield  {author} {\bibinfo {author} {\bibfnamefont {M.}~\bibnamefont {Nishioka}}\ and\ \bibinfo {author} {\bibfnamefont {H.}~\bibnamefont {Sato}},\ }\bibfield  {title} {\enquote {\bibinfo {title} {Mechanism of determination of the shedding frequency of vortices behind a cylinder at low reynolds numbers},}\ }\href@noop {} {\bibfield  {journal} {\bibinfo  {journal} {J. Fluid Mech.}\ }\textbf {\bibinfo {volume} {89}},\ \bibinfo {pages} {49--60} (\bibinfo {year} {1978})}\BibitemShut {NoStop}%
\bibitem [{\citenamefont {Kang}, \citenamefont {Choi},\ and\ \citenamefont {Lee}(1999)}]{kang1999laminar}%
  \BibitemOpen
  \bibfield  {author} {\bibinfo {author} {\bibfnamefont {S.}~\bibnamefont {Kang}}, \bibinfo {author} {\bibfnamefont {H.}~\bibnamefont {Choi}}, \ and\ \bibinfo {author} {\bibfnamefont {S.}~\bibnamefont {Lee}},\ }\bibfield  {title} {\enquote {\bibinfo {title} {Laminar flow past a rotating circular cylinder},}\ }\href@noop {} {\bibfield  {journal} {\bibinfo  {journal} {Phys. Fluids}\ }\textbf {\bibinfo {volume} {11}},\ \bibinfo {pages} {3312--3321} (\bibinfo {year} {1999})}\BibitemShut {NoStop}%
\bibitem [{\citenamefont {Tokumaru}\ and\ \citenamefont {Dimotakis}(1991)}]{tokumaru1991rotary}%
  \BibitemOpen
  \bibfield  {author} {\bibinfo {author} {\bibfnamefont {P.~T.}\ \bibnamefont {Tokumaru}}\ and\ \bibinfo {author} {\bibfnamefont {P.~E.}\ \bibnamefont {Dimotakis}},\ }\bibfield  {title} {\enquote {\bibinfo {title} {Rotary oscillation control of a cylinder wake},}\ }\href@noop {} {\bibfield  {journal} {\bibinfo  {journal} {J. Fluid Mech.}\ }\textbf {\bibinfo {volume} {224}},\ \bibinfo {pages} {77--90} (\bibinfo {year} {1991})}\BibitemShut {NoStop}%
\bibitem [{\citenamefont {Kumar}, \citenamefont {Cantu},\ and\ \citenamefont {Gonzalez}(2011)}]{kumar2011flow}%
  \BibitemOpen
  \bibfield  {author} {\bibinfo {author} {\bibfnamefont {S.}~\bibnamefont {Kumar}}, \bibinfo {author} {\bibfnamefont {C.}~\bibnamefont {Cantu}}, \ and\ \bibinfo {author} {\bibfnamefont {B.}~\bibnamefont {Gonzalez}},\ }\bibfield  {title} {\enquote {\bibinfo {title} {Flow past a rotating cylinder at low and high rotation rates},}\ }\href@noop {} {\bibfield  {journal} {\bibinfo  {journal} {J. Fluids Eng.}\ }\textbf {\bibinfo {volume} {133}},\ \bibinfo {pages} {041201} (\bibinfo {year} {2011})}\BibitemShut {NoStop}%
\bibitem [{\citenamefont {Darvishyadegari}\ and\ \citenamefont {Hassanzadeh}(2019)}]{darvishyadegari2019heat}%
  \BibitemOpen
  \bibfield  {author} {\bibinfo {author} {\bibfnamefont {M.}~\bibnamefont {Darvishyadegari}}\ and\ \bibinfo {author} {\bibfnamefont {R.}~\bibnamefont {Hassanzadeh}},\ }\bibfield  {title} {\enquote {\bibinfo {title} {Heat and fluid flow around two co-rotating cylinders in tandem arrangement},}\ }\href@noop {} {\bibfield  {journal} {\bibinfo  {journal} {International Journal of Thermal Sciences}\ }\textbf {\bibinfo {volume} {135}},\ \bibinfo {pages} {206--220} (\bibinfo {year} {2019})}\BibitemShut {NoStop}%
\bibitem [{\citenamefont {Van~Dijk}\ and\ \citenamefont {De~Lange}(2007)}]{van2007compressible}%
  \BibitemOpen
  \bibfield  {author} {\bibinfo {author} {\bibfnamefont {A.}~\bibnamefont {Van~Dijk}}\ and\ \bibinfo {author} {\bibfnamefont {H.}~\bibnamefont {De~Lange}},\ }\bibfield  {title} {\enquote {\bibinfo {title} {Compressible laminar flow around a wall-mounted cubic obstacle},}\ }\href@noop {} {\bibfield  {journal} {\bibinfo  {journal} {Computers \& Fluids}\ }\textbf {\bibinfo {volume} {36}},\ \bibinfo {pages} {949--960} (\bibinfo {year} {2007})}\BibitemShut {NoStop}%
\bibitem [{\citenamefont {Liu}\ \emph {et~al.}(2023)\citenamefont {Liu}, \citenamefont {Ding}, \citenamefont {Tao}, \citenamefont {Qu}, \citenamefont {Xie},\ and\ \citenamefont {Qiu}}]{liu2023numerical}%
  \BibitemOpen
  \bibfield  {author} {\bibinfo {author} {\bibfnamefont {Y.}~\bibnamefont {Liu}}, \bibinfo {author} {\bibfnamefont {Z.}~\bibnamefont {Ding}}, \bibinfo {author} {\bibfnamefont {Y.}~\bibnamefont {Tao}}, \bibinfo {author} {\bibfnamefont {J.}~\bibnamefont {Qu}}, \bibinfo {author} {\bibfnamefont {X.}~\bibnamefont {Xie}}, \ and\ \bibinfo {author} {\bibfnamefont {X.}~\bibnamefont {Qiu}},\ }\bibfield  {title} {\enquote {\bibinfo {title} {Numerical study on compressible flow around a circular cylinder in proximity to the wall},}\ }\href@noop {} {\bibfield  {journal} {\bibinfo  {journal} {Physics of Fluids}\ }\textbf {\bibinfo {volume} {35}} (\bibinfo {year} {2023})}\BibitemShut {NoStop}%
\bibitem [{\citenamefont {Xue}, \citenamefont {Li},\ and\ \citenamefont {Zhao}(2024)}]{xue2024compressibility}%
  \BibitemOpen
  \bibfield  {author} {\bibinfo {author} {\bibfnamefont {K.}~\bibnamefont {Xue}}, \bibinfo {author} {\bibfnamefont {Q.}~\bibnamefont {Li}}, \ and\ \bibinfo {author} {\bibfnamefont {L.}~\bibnamefont {Zhao}},\ }\bibfield  {title} {\enquote {\bibinfo {title} {Compressibility effect on flow characteristics over a circular cylinder at reynolds number of 3900},}\ }\href@noop {} {\bibfield  {journal} {\bibinfo  {journal} {Physics of Fluids}\ }\textbf {\bibinfo {volume} {36}} (\bibinfo {year} {2024})}\BibitemShut {NoStop}%
\bibitem [{\citenamefont {Biringen}\ and\ \citenamefont {Hatay}(1993)}]{biringen1993numerical}%
  \BibitemOpen
  \bibfield  {author} {\bibinfo {author} {\bibfnamefont {S.}~\bibnamefont {Biringen}}\ and\ \bibinfo {author} {\bibfnamefont {F.~F.}\ \bibnamefont {Hatay}},\ }\href@noop {} {\enquote {\bibinfo {title} {Numerical simulation of stability and stability control of high speed compressible rotating couette flow},}\ }\bibinfo {type} {Tech. Rep.}\ (\bibinfo {year} {1993})\BibitemShut {NoStop}%
\bibitem [{\citenamefont {Blackburn}\ and\ \citenamefont {Henderson}(1999)}]{blackburn1999study}%
  \BibitemOpen
  \bibfield  {author} {\bibinfo {author} {\bibfnamefont {H.~M.}\ \bibnamefont {Blackburn}}\ and\ \bibinfo {author} {\bibfnamefont {R.~D.}\ \bibnamefont {Henderson}},\ }\bibfield  {title} {\enquote {\bibinfo {title} {A study of two-dimensional flow past an oscillating cylinder},}\ }\href@noop {} {\bibfield  {journal} {\bibinfo  {journal} {J. Fluid Mech.}\ }\textbf {\bibinfo {volume} {385}},\ \bibinfo {pages} {255--286} (\bibinfo {year} {1999})}\BibitemShut {NoStop}%
\bibitem [{\citenamefont {Pralits}, \citenamefont {Giannetti},\ and\ \citenamefont {Brandt}(2013)}]{pralits2013three}%
  \BibitemOpen
  \bibfield  {author} {\bibinfo {author} {\bibfnamefont {J.~O.}\ \bibnamefont {Pralits}}, \bibinfo {author} {\bibfnamefont {F.}~\bibnamefont {Giannetti}}, \ and\ \bibinfo {author} {\bibfnamefont {L.}~\bibnamefont {Brandt}},\ }\bibfield  {title} {\enquote {\bibinfo {title} {Three-dimensional instability of the flow around a rotating circular cylinder},}\ }\href@noop {} {\bibfield  {journal} {\bibinfo  {journal} {J. Fluid Mech.}\ }\textbf {\bibinfo {volume} {730}},\ \bibinfo {pages} {5--18} (\bibinfo {year} {2013})}\BibitemShut {NoStop}%
\bibitem [{\citenamefont {Akoury}\ \emph {et~al.}(2008)\citenamefont {Akoury}, \citenamefont {Braza}, \citenamefont {Perrin}, \citenamefont {Harran},\ and\ \citenamefont {Hoarau}}]{el2008three}%
  \BibitemOpen
  \bibfield  {author} {\bibinfo {author} {\bibfnamefont {R.~E.}\ \bibnamefont {Akoury}}, \bibinfo {author} {\bibfnamefont {M.}~\bibnamefont {Braza}}, \bibinfo {author} {\bibfnamefont {R.}~\bibnamefont {Perrin}}, \bibinfo {author} {\bibfnamefont {G.}~\bibnamefont {Harran}}, \ and\ \bibinfo {author} {\bibfnamefont {Y.}~\bibnamefont {Hoarau}},\ }\bibfield  {title} {\enquote {\bibinfo {title} {The three-dimensional transition in the flow around a rotating cylinder},}\ }\href@noop {} {\bibfield  {journal} {\bibinfo  {journal} {J. Fluid Mech.}\ }\textbf {\bibinfo {volume} {607}},\ \bibinfo {pages} {1--11} (\bibinfo {year} {2008})}\BibitemShut {NoStop}%
\bibitem [{\citenamefont {Rao}\ \emph {et~al.}(2015)\citenamefont {Rao}, \citenamefont {Radi}, \citenamefont {Leontini}, \citenamefont {Thompson}, \citenamefont {Sheridan},\ and\ \citenamefont {Hourigan}}]{rao2015review}%
  \BibitemOpen
  \bibfield  {author} {\bibinfo {author} {\bibfnamefont {A.}~\bibnamefont {Rao}}, \bibinfo {author} {\bibfnamefont {A.}~\bibnamefont {Radi}}, \bibinfo {author} {\bibfnamefont {J.~S.}\ \bibnamefont {Leontini}}, \bibinfo {author} {\bibfnamefont {M.~C.}\ \bibnamefont {Thompson}}, \bibinfo {author} {\bibfnamefont {J.}~\bibnamefont {Sheridan}}, \ and\ \bibinfo {author} {\bibfnamefont {K.}~\bibnamefont {Hourigan}},\ }\bibfield  {title} {\enquote {\bibinfo {title} {A review of rotating cylinder wake transitions},}\ }\href@noop {} {\bibfield  {journal} {\bibinfo  {journal} {J. Fluids Struct.}\ }\textbf {\bibinfo {volume} {53}},\ \bibinfo {pages} {2--14} (\bibinfo {year} {2015})}\BibitemShut {NoStop}%
\bibitem [{\citenamefont {Theofilis}(2003)}]{theofilis2003advances}%
  \BibitemOpen
  \bibfield  {author} {\bibinfo {author} {\bibfnamefont {V.}~\bibnamefont {Theofilis}},\ }\bibfield  {title} {\enquote {\bibinfo {title} {Advances in global linear instability analysis of nonparallel and three-dimensional flows},}\ }\href@noop {} {\bibfield  {journal} {\bibinfo  {journal} {Prog. Aerosp. Sci.}\ }\textbf {\bibinfo {volume} {39}},\ \bibinfo {pages} {249--315} (\bibinfo {year} {2003})}\BibitemShut {NoStop}%
\bibitem [{\citenamefont {Pulliam}(1986)}]{pulliam1986implicit}%
  \BibitemOpen
  \bibfield  {author} {\bibinfo {author} {\bibfnamefont {T.~H.}\ \bibnamefont {Pulliam}},\ }\bibfield  {title} {\enquote {\bibinfo {title} {Implicit solution methods in computational fluid dynamics},}\ }\href@noop {} {\bibfield  {journal} {\bibinfo  {journal} {Appl. Numer. Math.}\ }\textbf {\bibinfo {volume} {2}},\ \bibinfo {pages} {441--474} (\bibinfo {year} {1986})}\BibitemShut {NoStop}%
\bibitem [{\citenamefont {Hoffmann}\ and\ \citenamefont {Chiang}(2000)}]{hoffmann2000computational}%
  \BibitemOpen
  \bibfield  {author} {\bibinfo {author} {\bibfnamefont {K.~A.}\ \bibnamefont {Hoffmann}}\ and\ \bibinfo {author} {\bibfnamefont {S.~T.}\ \bibnamefont {Chiang}},\ }\bibfield  {title} {\enquote {\bibinfo {title} {Computational fluid dynamics volume i},}\ }\href@noop {} {\bibfield  {journal} {\bibinfo  {journal} {Eng. Educ. Syst.}\ } (\bibinfo {year} {2000})}\BibitemShut {NoStop}%
\bibitem [{\citenamefont {Sagaut}\ \emph {et~al.}(2023)\citenamefont {Sagaut}, \citenamefont {Suman}, \citenamefont {Sundaram}, \citenamefont {Rajpoot}, \citenamefont {Bhumkar}, \citenamefont {Sengupta}, \citenamefont {Sengupta},\ and\ \citenamefont {Sengupta}}]{sagaut2023global}%
  \BibitemOpen
  \bibfield  {author} {\bibinfo {author} {\bibfnamefont {P.}~\bibnamefont {Sagaut}}, \bibinfo {author} {\bibfnamefont {V.~K.}\ \bibnamefont {Suman}}, \bibinfo {author} {\bibfnamefont {P.}~\bibnamefont {Sundaram}}, \bibinfo {author} {\bibfnamefont {M.~K.}\ \bibnamefont {Rajpoot}}, \bibinfo {author} {\bibfnamefont {Y.~G.}\ \bibnamefont {Bhumkar}}, \bibinfo {author} {\bibfnamefont {S.}~\bibnamefont {Sengupta}}, \bibinfo {author} {\bibfnamefont {A.}~\bibnamefont {Sengupta}}, \ and\ \bibinfo {author} {\bibfnamefont {T.~K.}\ \bibnamefont {Sengupta}},\ }\bibfield  {title} {\enquote {\bibinfo {title} {Global spectral analysis: Review of numerical methods},}\ }\href@noop {} {\bibfield  {journal} {\bibinfo  {journal} {Comput. Fluids}\ ,\ \bibinfo {pages} {105915}} (\bibinfo {year} {2023})}\BibitemShut {NoStop}%
\bibitem [{\citenamefont {Jameson}(2017)}]{jameson2017origins}%
  \BibitemOpen
  \bibfield  {author} {\bibinfo {author} {\bibfnamefont {A.}~\bibnamefont {Jameson}},\ }\bibfield  {title} {\enquote {\bibinfo {title} {Origins and further development of the jameson--schmidt--turkel scheme},}\ }\href@noop {} {\bibfield  {journal} {\bibinfo  {journal} {AIAA J.}\ }\textbf {\bibinfo {volume} {55}},\ \bibinfo {pages} {1487--1510} (\bibinfo {year} {2017})}\BibitemShut {NoStop}%
\bibitem [{\citenamefont {Sundaram}\ \emph {et~al.}(2023)\citenamefont {Sundaram}, \citenamefont {Sengupta}, \citenamefont {Suman},\ and\ \citenamefont {Sengupta}}]{sundaram2023non}%
  \BibitemOpen
  \bibfield  {author} {\bibinfo {author} {\bibfnamefont {P.}~\bibnamefont {Sundaram}}, \bibinfo {author} {\bibfnamefont {A.}~\bibnamefont {Sengupta}}, \bibinfo {author} {\bibfnamefont {V.~K.}\ \bibnamefont {Suman}}, \ and\ \bibinfo {author} {\bibfnamefont {T.~K.}\ \bibnamefont {Sengupta}},\ }\bibfield  {title} {\enquote {\bibinfo {title} {Non-overlapping high-accuracy parallel closure for compact schemes: Application in multiphysics and complex geometry},}\ }\href@noop {} {\bibfield  {journal} {\bibinfo  {journal} {ACM Trans. Parallel Comput.}\ }\textbf {\bibinfo {volume} {10}},\ \bibinfo {pages} {1--28} (\bibinfo {year} {2023})}\BibitemShut {NoStop}%
\bibitem [{\citenamefont {Sengupta}\ and\ \citenamefont {Tucker}(2020)}]{sengupta2020effects}%
  \BibitemOpen
  \bibfield  {author} {\bibinfo {author} {\bibfnamefont {A.}~\bibnamefont {Sengupta}}\ and\ \bibinfo {author} {\bibfnamefont {P.}~\bibnamefont {Tucker}},\ }\bibfield  {title} {\enquote {\bibinfo {title} {Effects of forced frequency oscillations and free stream turbulence on the separation-induced transition in pressure gradient dominated flows},}\ }\href@noop {} {\bibfield  {journal} {\bibinfo  {journal} {Physics of Fluids}\ }\textbf {\bibinfo {volume} {32}} (\bibinfo {year} {2020})}\BibitemShut {NoStop}%
\bibitem [{\citenamefont {Chew}, \citenamefont {Cheng},\ and\ \citenamefont {Luo}(1995)}]{chew1995numerical}%
  \BibitemOpen
  \bibfield  {author} {\bibinfo {author} {\bibfnamefont {Y.~T.}\ \bibnamefont {Chew}}, \bibinfo {author} {\bibfnamefont {M.}~\bibnamefont {Cheng}}, \ and\ \bibinfo {author} {\bibfnamefont {S.~C.}\ \bibnamefont {Luo}},\ }\bibfield  {title} {\enquote {\bibinfo {title} {A numerical study of flow past a rotating circular cylinder using a hybrid vortex scheme},}\ }\href {\doibase 10.1017/S0022112095003417} {\bibfield  {journal} {\bibinfo  {journal} {Journal of Fluid Mechanics}\ }\textbf {\bibinfo {volume} {299}},\ \bibinfo {pages} {35–71} (\bibinfo {year} {1995})}\BibitemShut {NoStop}%
\bibitem [{\citenamefont {Karabelas}\ \emph {et~al.}(2012)\citenamefont {Karabelas}, \citenamefont {Koumroglou}, \citenamefont {Argyropoulos},\ and\ \citenamefont {Markatos}}]{karabelas2012high}%
  \BibitemOpen
  \bibfield  {author} {\bibinfo {author} {\bibfnamefont {S.}~\bibnamefont {Karabelas}}, \bibinfo {author} {\bibfnamefont {B.}~\bibnamefont {Koumroglou}}, \bibinfo {author} {\bibfnamefont {C.}~\bibnamefont {Argyropoulos}}, \ and\ \bibinfo {author} {\bibfnamefont {N.}~\bibnamefont {Markatos}},\ }\bibfield  {title} {\enquote {\bibinfo {title} {High reynolds number turbulent flow past a rotating cylinder},}\ }\href@noop {} {\bibfield  {journal} {\bibinfo  {journal} {Applied mathematical modelling}\ }\textbf {\bibinfo {volume} {36}},\ \bibinfo {pages} {379--398} (\bibinfo {year} {2012})}\BibitemShut {NoStop}%
\bibitem [{\citenamefont {Aljure}\ \emph {et~al.}(2015)\citenamefont {Aljure}, \citenamefont {Rodr{\'\i}guez}, \citenamefont {Lehmkuhl}, \citenamefont {P{\'e}rez-Segarra},\ and\ \citenamefont {Oliva}}]{aljure2015influence}%
  \BibitemOpen
  \bibfield  {author} {\bibinfo {author} {\bibfnamefont {D.}~\bibnamefont {Aljure}}, \bibinfo {author} {\bibfnamefont {I.}~\bibnamefont {Rodr{\'\i}guez}}, \bibinfo {author} {\bibfnamefont {O.}~\bibnamefont {Lehmkuhl}}, \bibinfo {author} {\bibfnamefont {C.~D.}\ \bibnamefont {P{\'e}rez-Segarra}}, \ and\ \bibinfo {author} {\bibfnamefont {A.}~\bibnamefont {Oliva}},\ }\bibfield  {title} {\enquote {\bibinfo {title} {Influence of rotation on the flow over a cylinder at re= 5000},}\ }\href@noop {} {\bibfield  {journal} {\bibinfo  {journal} {International Journal of Heat and Fluid Flow}\ }\textbf {\bibinfo {volume} {55}},\ \bibinfo {pages} {76--90} (\bibinfo {year} {2015})}\BibitemShut {NoStop}%
\bibitem [{\citenamefont {Cheng}\ and\ \citenamefont {Luo}(2007)}]{cheng2007characteristics}%
  \BibitemOpen
  \bibfield  {author} {\bibinfo {author} {\bibfnamefont {M.}~\bibnamefont {Cheng}}\ and\ \bibinfo {author} {\bibfnamefont {L.-S.}\ \bibnamefont {Luo}},\ }\bibfield  {title} {\enquote {\bibinfo {title} {Characteristics of two-dimensional flow around a rotating circular cylinder near a plane wall},}\ }\href@noop {} {\bibfield  {journal} {\bibinfo  {journal} {Physics of Fluids}\ }\textbf {\bibinfo {volume} {19}} (\bibinfo {year} {2007})}\BibitemShut {NoStop}%
\bibitem [{\citenamefont {Theofilis}(2011)}]{theofilis2011global}%
  \BibitemOpen
  \bibfield  {author} {\bibinfo {author} {\bibfnamefont {V.}~\bibnamefont {Theofilis}},\ }\bibfield  {title} {\enquote {\bibinfo {title} {Global linear instability},}\ }\href@noop {} {\bibfield  {journal} {\bibinfo  {journal} {Ann. Rev. Fluid Mech.}\ }\textbf {\bibinfo {volume} {43}},\ \bibinfo {pages} {319--352} (\bibinfo {year} {2011})}\BibitemShut {NoStop}%
\bibitem [{\citenamefont {Sipp}\ and\ \citenamefont {Lebedev}(2007)}]{sipp2007global}%
  \BibitemOpen
  \bibfield  {author} {\bibinfo {author} {\bibfnamefont {D.}~\bibnamefont {Sipp}}\ and\ \bibinfo {author} {\bibfnamefont {A.}~\bibnamefont {Lebedev}},\ }\bibfield  {title} {\enquote {\bibinfo {title} {Global stability of base and mean flows: a general approach and its applications to cylinder and open cavity flows},}\ }\href@noop {} {\bibfield  {journal} {\bibinfo  {journal} {J. Fluid Mech.}\ }\textbf {\bibinfo {volume} {593}},\ \bibinfo {pages} {333--358} (\bibinfo {year} {2007})}\BibitemShut {NoStop}%
\bibitem [{\citenamefont {Sengupta}\ \emph {et~al.}(2018)\citenamefont {Sengupta}, \citenamefont {Suman}, \citenamefont {Sengupta},\ and\ \citenamefont {Bhaumik}}]{sengupta2018enstrophy}%
  \BibitemOpen
  \bibfield  {author} {\bibinfo {author} {\bibfnamefont {A.}~\bibnamefont {Sengupta}}, \bibinfo {author} {\bibfnamefont {V.}~\bibnamefont {Suman}}, \bibinfo {author} {\bibfnamefont {T.~K.}\ \bibnamefont {Sengupta}}, \ and\ \bibinfo {author} {\bibfnamefont {S.}~\bibnamefont {Bhaumik}},\ }\bibfield  {title} {\enquote {\bibinfo {title} {An enstrophy-based linear and nonlinear receptivity theory},}\ }\href@noop {} {\bibfield  {journal} {\bibinfo  {journal} {Physics of Fluids}\ }\textbf {\bibinfo {volume} {30}} (\bibinfo {year} {2018})}\BibitemShut {NoStop}%
\bibitem [{\citenamefont {Joshi}\ \emph {et~al.}(2025)\citenamefont {Joshi}, \citenamefont {Sengupta}, \citenamefont {Ajanif},\ and\ \citenamefont {Lestandi}}]{joshi2025comparing}%
  \BibitemOpen
  \bibfield  {author} {\bibinfo {author} {\bibfnamefont {B.}~\bibnamefont {Joshi}}, \bibinfo {author} {\bibfnamefont {A.}~\bibnamefont {Sengupta}}, \bibinfo {author} {\bibfnamefont {Y.}~\bibnamefont {Ajanif}}, \ and\ \bibinfo {author} {\bibfnamefont {L.}~\bibnamefont {Lestandi}},\ }\bibfield  {title} {\enquote {\bibinfo {title} {Comparing the highly-resolved onset of rayleigh--taylor and kelvin--helmholtz rayleigh--taylor instabilities},}\ }\href@noop {} {\bibfield  {journal} {\bibinfo  {journal} {European Journal of Mechanics-B/Fluids}\ ,\ \bibinfo {pages} {204382}} (\bibinfo {year} {2025})}\BibitemShut {NoStop}%
\bibitem [{\citenamefont {Joshi}\ \emph {et~al.}(2024)\citenamefont {Joshi}, \citenamefont {Sengupta}, \citenamefont {Sundaram},\ and\ \citenamefont {Sengupta}}]{joshi2024highly}%
  \BibitemOpen
  \bibfield  {author} {\bibinfo {author} {\bibfnamefont {B.}~\bibnamefont {Joshi}}, \bibinfo {author} {\bibfnamefont {T.~K.}\ \bibnamefont {Sengupta}}, \bibinfo {author} {\bibfnamefont {P.}~\bibnamefont {Sundaram}}, \ and\ \bibinfo {author} {\bibfnamefont {A.}~\bibnamefont {Sengupta}},\ }\bibfield  {title} {\enquote {\bibinfo {title} {Highly resolved peta-scale direct numerical simulations: Onset of kelvin--helmholtz rayleigh--taylor instability via pressure pulses},}\ }\href@noop {} {\bibfield  {journal} {\bibinfo  {journal} {Computers \& Fluids}\ }\textbf {\bibinfo {volume} {284}},\ \bibinfo {pages} {106442} (\bibinfo {year} {2024})}\BibitemShut {NoStop}%
\bibitem [{\citenamefont {Sengupta}\ and\ \citenamefont {Joshi}(2023)}]{sengupta2023effects}%
  \BibitemOpen
  \bibfield  {author} {\bibinfo {author} {\bibfnamefont {A.}~\bibnamefont {Sengupta}}\ and\ \bibinfo {author} {\bibfnamefont {B.}~\bibnamefont {Joshi}},\ }\bibfield  {title} {\enquote {\bibinfo {title} {Effects of stabilizing and destabilizing thermal gradients on reversed shear-stratified flows: Combined kelvin--helmholtz rayleigh--taylor instability},}\ }\href@noop {} {\bibfield  {journal} {\bibinfo  {journal} {Physics of Fluids}\ }\textbf {\bibinfo {volume} {35}} (\bibinfo {year} {2023})}\BibitemShut {NoStop}%
\bibitem [{\citenamefont {Sengupta}\ and\ \citenamefont {Sundaram}(2023)}]{sengupta2023compressibility}%
  \BibitemOpen
  \bibfield  {author} {\bibinfo {author} {\bibfnamefont {A.}~\bibnamefont {Sengupta}}\ and\ \bibinfo {author} {\bibfnamefont {P.}~\bibnamefont {Sundaram}},\ }\bibfield  {title} {\enquote {\bibinfo {title} {Compressibility effects on the flow past a t106a low-pressure turbine cascade},}\ }\href@noop {} {\bibfield  {journal} {\bibinfo  {journal} {Physics of Fluids}\ }\textbf {\bibinfo {volume} {35}} (\bibinfo {year} {2023})}\BibitemShut {NoStop}%
\bibitem [{\citenamefont {Sengupta}(2025)}]{sengupta2025compressible}%
  \BibitemOpen
  \bibfield  {author} {\bibinfo {author} {\bibfnamefont {A.}~\bibnamefont {Sengupta}},\ }\bibfield  {title} {\enquote {\bibinfo {title} {Compressible enstrophy transport for flow in a low-pressure turbine with unsteady wakes impinging at the inflow},}\ }in\ \href@noop {} {\emph {\bibinfo {booktitle} {Computational Fluid Dynamics: Novel Numerical and Computational Approaches: Methodology and Numerics}}}\ (\bibinfo  {publisher} {Springer},\ \bibinfo {year} {2025})\ pp.\ \bibinfo {pages} {59--85}\BibitemShut {NoStop}%
\bibitem [{\citenamefont {San}\ and\ \citenamefont {Maulik}(2018)}]{san2018machine}%
  \BibitemOpen
  \bibfield  {author} {\bibinfo {author} {\bibfnamefont {O.}~\bibnamefont {San}}\ and\ \bibinfo {author} {\bibfnamefont {R.}~\bibnamefont {Maulik}},\ }\bibfield  {title} {\enquote {\bibinfo {title} {Machine learning closures for model order reduction of thermal fluids},}\ }\href@noop {} {\bibfield  {journal} {\bibinfo  {journal} {Applied Mathematical Modelling}\ }\textbf {\bibinfo {volume} {60}},\ \bibinfo {pages} {681--710} (\bibinfo {year} {2018})}\BibitemShut {NoStop}%
\bibitem [{\citenamefont {Bengio}\ \emph {et~al.}(2017)\citenamefont {Bengio}, \citenamefont {Goodfellow}, \citenamefont {Courville} \emph {et~al.}}]{bengio2017deep}%
  \BibitemOpen
  \bibfield  {author} {\bibinfo {author} {\bibfnamefont {Y.}~\bibnamefont {Bengio}}, \bibinfo {author} {\bibfnamefont {I.}~\bibnamefont {Goodfellow}}, \bibinfo {author} {\bibfnamefont {A.}~\bibnamefont {Courville}},  \emph {et~al.},\ }\href@noop {} {\emph {\bibinfo {title} {Deep learning}}}\ (\bibinfo  {publisher} {MIT Press Cambridge, MA},\ \bibinfo {year} {2017})\BibitemShut {NoStop}%
\end{thebibliography}%


\end{document}